\documentclass[12pt, draftclsnofoot, onecolumn]{IEEEtran}
\ifCLASSINFOpdf

\else

\fi
\usepackage[cmex10]{amsmath}
\usepackage{amssymb}
\usepackage{cite}
\usepackage{graphicx}
\usepackage{array,color}
\usepackage{amsmath}
\usepackage{stfloats}
\usepackage{graphicx}
\usepackage{subfigure}
\usepackage{tabularx}
\usepackage{epsfig,epsf,color,balance,cite}
\usepackage{algorithmic}
\usepackage{algorithm}
\usepackage{bm}
\begin{document}
%
\title{Totally Distributed Energy-Efficient Transmission in MIMO Interference Channels}


\author{Cunhua Pan, Wei Xu, \IEEEmembership{Senior Member, IEEE}, Jiangzhou Wang, \IEEEmembership{Senior Member, IEEE}, Hong Ren, Wence Zhang, Nuo Huang, and Ming Chen
\thanks{This work was supported by National 863 High Technology Development Project (No. 2014AA01A701),  National Nature Science Foundation of China (Nos. 61172077 \& 61372106 \& 61223001 \& 61471114), the Scientific Research Foundation of Graduate School of Southeast University (No. YBJJ1433).
C. Pan, W. Xu, H. Ren, W. Zhang, N. Huang and M. Chen are with National Mobile Communications Research Laboratory, Southeast University, Nanjing 210096, China. (Email:\{cunhuapan, wxu, renhong, wencezhang, huangnuo, chenming\}@seu.edu.cn).
J. Wang is with the School of Engineering and Digital Arts, University of Kent, Canterbury, Kent, CT2 7NZ, U.K. (Email:{j.z.wang@kent.ac.uk}).
Part of this work was presented in IEEE Globecom 2014.}
}

\maketitle



\vspace{-0.6cm} \begin{abstract}
In this paper, we consider the problem of maximizing the energy efficiency (EE) for  multi-input multi-output (MIMO) interference channels, subject to the per-link power constraint. To avoid extensive information exchange among all links, the optimization problem is formulated as a noncooperative game, where each link maximizes its own EE. We show that this game always admits a Nash equilibrium (NE) and the sufficient condition for the uniqueness of the NE is derived for the case of large enough maximum transmit power constraint. To reach the NE of this game, we develop a totally distributed EE algorithm, in which each link  updates its own transmit covariance matrix in a completely distributed and asynchronous way: Some players may update their solutions more frequently than others or even use the outdated interference information. The sufficient conditions that guarantee the global convergence of the proposed algorithm to the NE of the game have been given as well. We also study the impact of the circuit power consumption on the sum-EE performance of the proposed algorithm in the case when the links are separated sufficiently far away. Moreover, the tradeoff between the sum-EE and the sum-spectral efficiency (SE) is investigated with the proposed algorithm under two special cases: 1) low transmit power constraint regime; 2) high transmit power constraint regime.  Finally, extensive simulations are conducted to evaluate the impact of various system parameters on the system performance.

\end{abstract}



%
\IEEEpeerreviewmaketitle


\section{Introduction}
Past few years have witnessed tremendous advancement in wireless communications, including the significant improvement of transmission rate \cite{Huiling-2009,Huiling-2012}. However, the impact of power consumption on the environment is neglected. It is reported that  the total energy consumption of the communications takes up more than 3 percent of the worldwide electric energy consumption \cite{Fettweis-2008} and the portion is expected to increase due to the explosive growth of high-data-rate applications in the future. Hence, energy efficiency (EE) has gained lots of attention and will be one of key issues in future fifth-generation (5G) mobile networks \cite{Andrews-2014}. On the other hand, the interference channel (IC) has been modeled mathematically for many practical systems where multiple uncoordinated links share the same channel, such as femtocells, ad hoc wireless networks, cognitive radio, etc \cite{Andrews-2014,Chao2007}. Furthermore, due to the development of advanced multi-antenna techniques \cite{Kuo-2011,Sai-2014}, each transmission node is able to accommodate multiple antennas \cite{Vaze2012,shan2014}. It is well known that multi-input multi-output (MIMO) system has the great potential for providing high SE by employing spatial multiplexing techniques \cite{Gesbert-2007}. Hence, it is of great importance to study the energy efficient transmission strategy in MIMO ICs.

This paper focuses on the EE maximization problem for MIMO ICs with per-link transmit power constraint. To solve this problem, one may consider centralized solutions, which require a central processing unit (CPU) to collect all complex-valued channel matrices over the network. The CPU will compute  all links' transmit covariance matrices and send them to the corresponding links. Hence, for large-scale networks, the centralized approaches suffer from heavy feedback overhead and high computational complexity, which hinders  practical implementations. Moreover, there may not exist a CPU for some wireless networks, such as ad hoc or wireless sensor networks.

Recently, distributed algorithms to deal with this problem attract intensive attentions \cite{Palomar-2006,Yanqing-2014,kim2011optimal,Qingjiang-2011}. Here, ``distributed'' means that precoders can be computed at the transmitters with only local channel knowledge and limited (or no) information exchange over different links. Generally, distributed processing for MIMO systems has the benefits of low communication exchange overhead, low computational complexity,  more scalability, low system costs, etc. The classical distributed algorithm based on dual decomposition technique is designed to decompose the coupling constraints among links \cite{Palomar-2006,Yanqing-2014}. Specifically, by introducing the dual variables associated with the coupling constraints, the original problem can be divided into several independent subproblems, each of which can be solved in a distributed way. Then, all the links exchange some necessary information to update the dual variables.
For more practical networks with individual link power budget constraints,  \cite{ChenziJiang-2013} devised a  decentralized beamforming EE (DBFEE) algorithm for symmetric MIMO ICs, where the distance from one transmitter to its desired receiver is identical for all links, and each transmitter has the same distance to all its unintended receivers. In each iteration, all the receivers should feed back the equivalent channel matrices to all the transmitters in the network.  In \cite{ShiwenTSP}, the authors designed a two-layer EE (TLEE) algorithm based on the generalized weighted minimum mean square error (WMMSE) approach \cite{Qingjiang-2011}: The inner layer to update precoders/decoders; the outer layer to update some parameters. Similar to the algorithm in \cite{ChenziJiang-2013},  in each inner iteration each receiver should feed back the updated weight matrix and the positive definite covariance matrix to all the transmitters in the network. However, for these distributed algorithms, in each iteration each receiver needs to compute the necessary complex-valued matrix and then feed it back to all the transmitters in the network, which could induce serious implementation challenges such as a large amount of feedback overhead, poor scalability and heavy computational burden at the receivers. Moreover, all the links should be synchronous, which is difficult to be satisfied, especially for large-scale ad hoc networks or wireless sensor networks. One novel distributed algorithm based on the adaptive price  was proposed in \cite{Pan2015} to deal with the weighted sum EE maximization problem for single-input single-output (SISO) ICs.

Hence, one distributed algorithm with much lower feedback overhead is more desirable. Noncooperative game theoretical approach  has been recognized as a powerful tool to devise totally distributed algorithms, in which each link just maximizes its utility  without the need of information exchanges among the links. A number of researches have applied game theory to design energy efficient communications for  ICs \cite{Meshkati-2007,Betz-2008,Miao-2011,Zappone-2013} or multiple access channel (MAC) \cite{buzzi2008energy}. The EE optimization problem in flat fading single-input single-output (SISO) ICs was considered in \cite{Betz-2008}, where one distributed algorithm  based on non-cooperative game was proposed. Both the existence and uniqueness of Nash equilibrium (NE) were analyzed. This work was extended to frequency-selective channel in \cite{Miao-2011} and to a relay channel in \cite{Zappone-2013}. Although \cite{Miao-2011} proved the existence of the NE, regarding the uniqueness of the NE, \cite{Miao-2011} only showed that the number of NEs is determined by the cross-channel gains and the direct channel gains, without quantifying how they are related with each other.  All these studies considered single-antenna ICs and they apply for multi-antenna case only if the transmit powers are optimized with fixed transmit directions as shown in \cite{Zappone-2014}. In \cite{buzzi2008energy}, the authors considered the EE maximization problem for the MIMO uplink systems with each user transmitting only one stream. This problem is formulated as a non-cooperative game, where the uniqueness of NE is guaranteed by the fact that the EE function is S-shaped.

In this paper, we apply the non-cooperative game theoretical approach to deal with the EE
maximization problem for MIMO ICs, where each link attempts to maximize its own EE by
jointly optimizing transmit power and beamformers. It is a nontrivial extension of the SISO case in \cite{Miao-2011}. There is an explicit relationship between the power allocation among different links and the achievable rates in the SISO case. This property is critical in deriving the conditions of the existence and uniqueness of the NE by using the standard function \cite{Miao-2011}. However, in MIMO systems, this relationship is implicit, as power allocation is carried out through matrix manipulations. Moreover especially when multiplexing is utilized with the MIMO,  different from the beamforming case in \cite{buzzi2008energy} for MAC,  the EE design is shown more general for the MIMO IC and hence more difficult since there exist both inter-node (mutual) interference and intra-node inference in the MIMO IC using multiplexing.

\subsection{Related Work}

Recently, distributed algorithms for MIMO ICs have been extensively studied in the literature, such as linear iterative approximation (LIA) algorithm \cite{kim2011optimal}, the WMMSE algorithm \cite{Qingjiang-2011}, noncooperative game theoretic algorithm \cite{Sigen-2003},  etc. For the LIA algorithm, it is designed based on the first-order Taylor expansion of the non-convex part of the weighted sum spectrum efficiency (SE) objective function. Sequential convex optimization approaches were then presented for dealing with various scenarios, e.g.,  the multi-band scenario in \cite{nguyen2012price}, the MAC in \cite{Nguyen2014MAC},  the broadcast channel in \cite{Nguyen2014BC}, the cognitive radio networks in \cite{Yanqing-2014}. However, in the LIA algorithm only one user is allowed to update its covariance matrix at one time, which may lead to significant latency especially in dense networks. By establishing the equivalence between the weighted sum SE problem and weighted sum mean square error minimization problem, \cite{Qingjiang-2011} proposed the WMMSE algorithm that allows multiple users to update simultaneously. In this algorithm, the local optimal solution is obtained via alternatively optimizing the linear transceivers and iteratively updating the weight matrices. The authors show that when the utility function satisfies some conditions, the algorithm is guaranteed to converge to the stationary point of the original problem. The WMMSE algorithm has been applied in various setups, please see \cite{Baligh2014SPM} and references therein.

However, to successfully implement the ILA and WMMSE algorithms in a distributed manner,  two assumptions are required: 1) Perfect channel reciprocal between the uplink phase and the downlink phase (in time-division duplexing mode); 2) Synchronization between all the links. In practice, the communication systems usually operate in frequency-division duplexing (FDD) mode. The channel reciprocal is thus hard to achieve. In addition, for wireless senor or ad hoc networks, asynchronous among the links is more desirable. More importantly, in each iteration each receiver needs to calculate the pricing matrix in the ILA algorithm or the weight matrix in the WMMSE algorithm, and then feeds them back  to all the transmitters in the network, which posses serious implementation issues such as a large amount of heavy feedback overhead, poor scalability and heavy computational burden at the receivers.

On the other hand, non-cooperative game theoretical approaches have attracted extensive attentions. Studies in this direction are plentiful in literature, e.g., \cite{lasaulce2007power} for the MIMO  MAC, \cite{Sigen-2003,scutari2008competitive,Scutari-2009,scutari2010mimo,wang2011robust,Nguyen2011TSP,Nguyen2014BD,zhou2014network} for the MIMO ICs. In \cite{lasaulce2007power}, the authors formulated the problem as a non-cooperative game and the authors proved that each user's optimal eigenvectors do not depend on the channels of others. Based on this fact, the authors showed that the existence and uniqueness of NE is guaranteed when the numbers of transmit and receive antennas become large. The main technique in \cite{lasaulce2007power} is  random matrix theory.  For the MIMO ICs, the authors in \cite{Sigen-2003}  first utilized the noncooperative game framework to deal with the SE maximization problem, where the iterative water-filling algorithm was proposed to find the NE of the game. However, the existence of the NE was only shown by the simulation results without theoretic guarantees, neither the NE uniqueness. Then, in \cite{scutari2008competitive}, Scutati \emph{et al} proved the existence of NE and provided sufficient conditions for the uniqueness of the NE, which can be checked in practice. However, the results are only valid for square nonsingular channel matrices. Later on, they generalized the results to a more general case with arbitrary channel dimensions in \cite{Scutari-2009}.  The cognitive radio network with null shaping constraints on the primary user is considered in \cite{scutari2010mimo} and its robust version in \cite{wang2011robust}. Most recently, these works were extended to the multicell case in \cite{Nguyen2011TSP,Nguyen2014BD} with multiple users per cell. In \cite{zhou2014network}, the authors formulated the SE maximization problem as a cooperative game. Specifically, by fixing the outgoing cooperative set and incoming cooperative set, the authors formulated this problem as a non-cooperative game, where the existence and uniqueness of the NE was analyzed. Then, the coalitional game theory \cite{saad2009coalitional} was applied to obtain the stable of the cooperative set. The work that is most closely related to ours is \cite{Scutari-2009}, where the noncooperative game was formulated for the SE maximization problem of the MIMO ICs and one asynchronous distributed algorithm was proposed to reach the NE of the game.

In contrast to the most of the above cited papers which focus on the (weighted) sum SE problem, in this work we consider the EE maximization problem. For SE optimization problems, it is known that all the transmitters use full power during transmission in order to maximize its own SE. Based on this fact, the best response strategy at the NE can be written in a closed-form water-filling solution, which can be interpreted as a projector on the convex and closed set. This interpretation enables the authors to derive the uniqueness of the game's NE \cite{Scutari-2009}. However, the study of  EE maximization problem cannot be obtained by employing the methodologies  since the transmitters in fact use a portion of the power, instead of full, to achieve energy efficient transmission.

\subsection{Contributions}

In this paper, we apply the non-cooperative game theoretical approach to deal with the EE maximization problem for MIMO ICs, where each link attempts to maximize its own EE by jointly optimizing transmit power and beamformers.

The main contributions and observations of our work are summarized as follows.
\begin{enumerate}
  \item The EE maximization problem in MIMO interference channels is modeled as a noncooperative game where each MIMO link  competes against the others by choosing its transmit covariance matrix to maximize its own EE. We show that the NE of this game always exists and derive sufficient conditions for the uniqueness of the NE for the case of large enough maximum transmit power constraint.
  \item To reach the NE of the game, we provide a totally distributed EE algorithm named Asynchronous Distributed Energy-Efficient (ADEE) algorithm, which is the extended version of simultaneous updating proposed in \cite{Miao-2011}. In this algorithm, all users apply the fractional programming to update the transmit covariance matrices and these updates can be performed in a totally asynchronous way, which means some links may update their transmit covariance matrices more frequently than the others and they may even use the outdated information of the measurement of the interference generated by the other links. In addition, during the updating procedure of the algorithm, there is no need for the links to exchange the signaling overhead mutually. These features make our distributed algorithm more appealing for practical implementations. We provide the  sufficient conditions for the global convergence of this algorithm to the unique NE of the game. Interestingly, we find that these conditions coincide with the  conditions for the uniqueness of NE.
  \item We study the impact of the circuit power consumption on the overall SE and EE performance of the system for one special case when the links are separated sufficiently far away. We show that the overall SE  increases with the circuit power consumption, but the overall EE decreases with it. Although this trend is derived for this special case, from simulations we find the trend holds for the general case when the interference among the links is sizeable.
      This observation implies that when the circuit power consumption increases, we should enhance the transmit rate or SE in order to obtain the best EE performance.
  \item The tradeoff between SE and EE is investigated for the proposed algorithm (denoted as ADEE algorithm) and the SE maximization algorithm (denoted as  ADSE algorithm) in \cite{Scutari-2009}. Two special cases are studied: the transmit power constraint approaches zero or infinity. For the case of low transmit power constraint, we show that both algorithms use all power to transmit and thus achieve the same performance in terms of the overall SE and EE performance. However, for the latter case, the ADSE algorithm always uses all available power to transmit, yielding severe inference over the network. In this case, the SE achieved by the ADSE algorithm will not increase. Then, the EE achieved by the ADSE algorithm will approach zero due to the significant power consumption. On the other hand, for the EE metric, the ADEE algorithm is unwilling to consume all power in this case. As a result, the SE and EE achieved by the ADEE algorithm will become constant in the case of the high transmit power constraint.
\end{enumerate}

The rest of the paper is organized as follows. In Section \ref{systemmodel}, we introduce the system model and formulate the optimization problem as a strategic noncooperative game. Then, we show that this game always admits a NE and derive sufficient conditions for the uniqueness of the NE in Section \ref{existence}. To reach the NE, a totally asynchronous and distributed algorithm is given in Section \ref{algorithm}. In Section \ref{performanceana}, we study the impact of the circuit power consumption on the system performance in terms of the sum-SE and sum-EE, along with the study for the tradeoff between the sum-EE and sum-SE for the proposed algorithm. Section \ref{simulation} provides representative numerical results to study the effects of different system parameters on the proposed algorithm. Finally, some conclusions are drawn in Section \ref{conclusion}.

Notations:  ${( \cdot )^{\rm{*}}}$, ${( \cdot )^T}$, ${( \cdot )^H}$,  ${\rm{vec}}(\cdot)$, ${\rm{E}}\{\cdot \} $, ${\rm{tr()}}$ and $ \otimes $ are conjugate, transpose, Hermitian transpose,  stacking vectorization operator, expectation operator, trace operators and the Kronecker product operator \cite{Horn-2012}, respectively. Uppercase and lowercase boldface denote matrices and vectors, respectively. For matrix ${\bf{A}}$,  ${\left[ {\bf{A}} \right]_{:,k}}$ and ${[{\bf{A}}]_{i,j}}$ represent the $k^{\rm{th}}$ column of  ${\bf{A}}$ and the $(i,j)$  element of matrix ${\bf{A}}$, respectively. ${\left\| {\bf{A}} \right\|_2}$ denotes the spectral norm of  ${\bf{A}}$  \cite{Horn-2012}. ${\left\| {\bf{A}} \right\|_F}$ denotes the Frobenius norm of   ${\bf{A}}$.
${\lambda _{\min }}({\bf{A}})$ stands for the minimum eigenvalue of ${\bf{A}}$.
${\bf{A}}\succeq{\bf{B}}$ means ${\bf{A}} - {\bf{B}}$ is positive semidefinite. The spectral radius of  ${\bf{A}}$ is denoted by $\rho {\rm{(}}{\bf{A}})$ \cite{Horn-2012}. ${\rm{rank}}({\bf{A}})$ denotes the rank of  ${\bf{A}}$. For vector ${\bf{a}} \in {\mathbb{C}^{n \times 1}}$, ${\left\| {\bf{a}} \right\|_2}$ represents the Euclidean norm defined as ${\left\| {\bf{a}} \right\|_2} = \sqrt {{{\bf{a}}^H}{\bf{a}}} $.  The sets of $m \times n$ complex matrices,  $n \times n$ complex positive semidefinite and definite matrices are denoted by ${\mathbb{C}^{m \times n}}$, $\mathbb{S}_ + ^{n \times n}$ and $\mathbb{S}_ {++} ^{n \times n}$, respectively. ${{\bf{D}}_{\bf{X}}}{\bf{Y}}$ denotes the Jacobian matrix of function ${\bf{Y}}$ with respect to (w.r.t.) ${\bf{X}}$ \cite{Hjorungnes-2007}. ${\bf{I}}$ and ${\bf{0}}$ represent the identity and zero matrices with appropriate dimensions, respectively. ${\left[ x \right]^ + }$  is equivalent to  $\max \left\{ {0,x} \right\}$.

\section{System Model and Problem Formulation}\label{systemmodel}

Consider a $K$-link MIMO interference channel with $K$ transmitter-receiver  pairs. Each link consists of one transmitter with $M$ transmit antennas and one receiver with $N$ receiver antennas. All links are simultaneously communicating over the same channel. At receiver $k$ \footnote{``Receiver $k$'' represents the receiver of the $k^{\rm{th}}$ link. In the following, ``transmitter $k$'' means the transmitter of the $k^{\rm{th}}$ link.}, the received complex baseband signal vector ${{\bf{y}}_k} \in{\mathbb{C}^{N\times 1}}$ is given by
\begin{equation}\label{receivesingal}
    {{\bf{y}}_k} = {{\bf{H}}_{k,k}}{{\bf{x}}_k} + \sum\limits_{j \ne k} {{{\bf{H}}_{jk}}{{\bf{x}}_j}}  + {{\bf{n}}_k},
\end{equation}
where ${{\bf{x}}_k} \in {\mathbb{C}^{M\times 1}}$  denotes the transmit signal vector of  link $k$, ${{\bf{H}}_{kk}} \in {\mathbb{C}^{N \times M}}$ is the direct channel matrix of link $k$, ${{\bf{H}}_{jk}} \in {\mathbb{C}^{N \times M}}$ is the cross-channel matrix from  transmitter $j$ to receiver $k$, and ${{\bf{n}}_k} \in \mathbb{C}^{N\times 1}$  is circularly symmetric, zero-mean, complex Gaussian noise with normalized identity covariance matrix. For each link, the total average transmit power should satisfy the per-link power constraint:
\begin{equation}\label{powerconstraint}
    {P_k} = {\rm{tr(}}{{\bf{Q}}_k}{\rm{)}} \le {P_{\rm{T}}},
\end{equation}
where ${{\bf{Q}}_k} = E\left\{ {{{\bf{x}}_k}{\bf{x}}_k^H} \right\}$  is the covariance matrix of ${{\bf{x}}_k}$ and ${P_{\rm{T}}}$ is the maximum transmit power.

To reduce the complexity of decoding at the receivers, it is assumed that joint decoding of the interfering signals is not an option and the interference is treated as noise at the receivers. Thus, the SE of link $k$ is given by (in bit/s/Hz)
\begin{equation}\label{rate}
    {C_k}({{\bf{Q}}_k},{{\bf{Q}}_{ - k}}) = \log_2 |{\bf{I}} + {\bf{H}}_{k,k}^H{\bf{R}}_k^{ - 1}{{\bf{H}}_{k,k}}{{\bf{Q}}_k}|,
\end{equation}
where ${{\bf{R}}_k} \buildrel \Delta \over = {\bf{I}} + \sum\nolimits_{j \ne k} {{{\bf{H}}_{j,k}}{{\bf{Q}}_j}{\bf{H}}_{j,k}^H} $ represents the interference-plus-noise (IPN) covariance matrix at  receiver $k$ and ${{\bf{Q}}_{ - k}}\buildrel \Delta \over = [{{\bf{Q}}_1}, \cdots ,{{\bf{Q}}_{k - 1}},{{\bf{Q}}_{k + 1}}, \cdots ,{{\bf{Q}}_K}]$ denotes the set of all links' covariance matrices, except that of link $k$. In this work, it is assumed that each receiver $k$ can perfectly measure the IPN covariance matrix ${{\bf{R}}_k}$ and estimate the direct channel ${\bf{H}}_{k,k}$, and then report them back to transmitter $k$. The channels are assumed to vary sufficiently slowly such that it can be considered fixed during the transmission.

In order to design energy efficient transmissions, the total power consumption should be considered at each link $k$: power used for reliable data transmission ${P_k}$, circuit power during transmission $P_{\rm{C}}$, which is
the  power consumed by the mixers, filters and digital-to-analog converters, digital signal processing (DSP), etc.  The power consumption of the DSP depends on the number of computations of the algorithm and the signaling overhead. It is difficult to accurately model this kind of power consumption. Thus, the circuit power $P_{\rm{C}}$ is modeled as a constant in this work for the sake of analysis, as simplified in most of the existing works \cite{Betz-2008,Miao-2011,ChenziJiang-2013,Zappone-2013,Zappone-2014}. Even though, we will evaluate the system performance under different $P_{\rm{C}}$  via simulations. Then, the EE  (in bits/Hz/Joule) of  link $k$, defined as the ratio of SE to the total power consumption, is given by
\begin{equation}\label{EE}
    {\rm{EE}}_k({{\bf{Q}}_k},{{\bf{Q}}_{ - k}}) = \frac{{{C_k}({{\bf{Q}}_k},{{\bf{Q}}_{ - k}})}}{{{P_k}{\rm{ + }}{P_{\rm{C}}}}}.
\end{equation}

Since our goal is to devise a totally distributed algorithm that requires neither a CPU nor information exchange among the links, we formulate the optimization problem as the following noncooperative game:
\begin{equation}\label{game}
    \begin{array}{*{20}{c}}
{({\cal G}):}&\begin{array}{l}
\mathop {\max }\limits_{{{\bf{Q}}_k}} \ {\rm{EE}}_k({{\bf{Q}}_k},{{\bf{Q}}_{ - k}}),\\
{\rm{s}}{\rm{.t}}{\rm{. }}\quad {{\bf{Q}}_k} \in {{\cal W}_k}
\end{array}&{\forall k \in {\Psi _K}},
\end{array}
\end{equation}
where  ${\rm{EE}}_k({{\bf{Q}}_k},{{\bf{Q}}_{ - k}})$ is the payoff function of link  $k$  defined in (\ref{EE}), ${\Psi _K} \buildrel \Delta \over = \left\{ {1,2, \cdots ,K} \right\}$ is the set of all links and ${{\cal W}_k}$  is the set of admissible strategies of link $k$, defined as
\begin{equation}\label{admissibleset}
    {{\cal W}_k} = \{ {{\bf{Q}}_k} \in {\mathbb{C}^{M \times M}}|{{\bf{Q}}_k} \in \mathbb{S}_ + ^{M \times M},{\rm{tr}}({{\bf{Q}}_k}) \le {P_{\rm{T}}}\}.
\end{equation}

In this game, each user competes against the others by choosing his own covariance matrix that maximizes his own payoff function subject to the strategy set. A solution of the game to reach a NE is when each link, given the strategic profiles of the others, does not get any increase in its objective  by unilaterally changing its own strategy and is formally defined as follows.

\itshape \textbf{Definition 1:}  \upshape A  strategic profile  ${{\bf{Q}}^\star} = {({\bf{Q}}_k^\star)_{k \in {\Psi _K}}} \in {{\cal W}_1} \times  \cdots  \times {{\cal W}_K}$ is a NE of game  ${\cal G}$ if
\begin{equation}\label{NE}
    {\rm{EE}}_k({\bf{Q}}_k^\star,{\bf{Q}}_{ - k}^\star)\! \ge\! {\rm{EE}}_k({{\bf{Q}}_k},{\bf{Q}}_{ - k}^\star),\forall {{\bf{Q}}_k} \in {{\cal W}_k},\forall k \in {\Psi _K}.
\end{equation}

In the forthcoming sections, we first show that game ${\cal G}$  always admits at least one NE. In general, game ${\cal G}$  may admit multiple NEs, depending on the level of the interference from the other links \cite{Miao-2011}. Then, we study the uniqueness condition of the NE and provide a totally distributed algorithm to reach such a NE.

\section{Existence and Uniqueness of the NE}\label{existence}

\subsection{Existence of NE}

Whether NE exists depends largely on the properties of the payoff function. In the sequel, we first study the property of the EE function and then check the existence of NE.

\itshape \textbf{Lemma 1:}  \upshape Given the other links' strategy ${{\bf{Q}}_{ - k}}$, the EE function of link $k$, i.e., ${\rm{EE}}_k({{\bf{Q}}_k},{{\bf{Q}}_{ - k}})$,  is quasiconcave in  ${{\bf{Q}}_k}$. Furthermore, if the channel matrix ${{{\bf{H}}_{k,k}}}$ is full column rank,  ${\rm{EE}}_k({{\bf{Q}}_k},{{\bf{Q}}_{ - k}})$ is strictly quasiconcave in  ${{\bf{Q}}_k}$\\
\itshape \textbf{Proof:}   \upshape Please see Appendix \ref{prooflemma1}.   \hfill $\Box$

Based on Lemma 1, the existence of NE is given in the following theorem.

\itshape \textbf{Theorem 1:}  \upshape Game ${\cal G}$ always admits at least one NE for any set of channels and transmit power constraints.

\itshape \textbf{Proof:}   \upshape Please see Appendix \ref{proofTheorem1}.   \hfill $\Box$

\subsection{Uniqueness of NE}
In this subsection, we first study the uniqueness condition for the case of full column-rank channel matrices. Then, we extend the results to the more general case without making any restrictive assumptions on the channel structure.

1) \emph{Case of full column-rank channel matrices}

For the SE maximization game in \cite{Scutari-2009}, all the transmitters use full power during transmission in order to maximize its own SE. Based on this fact, the best response strategy at the NE can be written in a closed-form water-filling solution, which can be interpreted as a projector on the convex and closed set. This interpretation enables the authors to derive the uniqueness condition of the game's NE. However, the study of our EE maximization problem cannot be obtained by employing the methodologies developed in \cite{Scutari-2009} since the transmitters do not always, and in fact not in most cases, use the full power to transmit in our EE case. Here, we consider the case when the maximum transmit power is large enough so that each transmitter only uses portion of the maximum power at the NE.

Before providing the uniqueness condition for the EE maximization game, we first introduce some useful intermediate results. Given the other links' strategy ${{\bf{ Q}}_{ - k}}$, the best response of the transmit covariance matrix of link $k$ is denoted as
\begin{equation}\label{mappingfunction}
    {{\bf{F}}_k}({{\bf{ Q}}_{ - k}}) = \arg \mathop {\max }\limits_{{{\bf{ Q}}_k} \in { {\cal W}_k}} {\rm{EE}}_k({{\bf{ Q}}_k},{{\bf{Q}}_{ - k}})
\end{equation}
where  function ${{\bf{F}}_k}({{\bf{ Q}}_{ - k}})$: $ { {\cal W}_{ - k}}\mapsto {\mathbb{C}^{M \times M}}$  is a complex matrix-valued function with ${{\cal W}_{ - k}} \buildrel \Delta \over = {{\cal W}_1} \times  \cdots  \times {{\cal W}_{k - 1}} \times {{\cal W}_{k + 1}} \times  \cdots  \times {{\cal W}_K}$. We introduce the following mapping function,
\begin{equation}\label{overallmapping}
    {\bf{F}}({\bf{ Q}}) = {{\bf{F}}_1}({{{\bf{ Q}}}_{ - 1}}) \times  \cdots  \times {{\bf{F}}_K}({{{\bf{ Q}}}_{ - K}}):{\cal W} \mapsto {\cal W}
\end{equation}
where ${\cal W} \buildrel \Delta \over = {{\cal W}_1} \times  \cdots  \times {{\cal W}_K}$ and ${\bf{Q}} \buildrel \Delta \over = [{{\bf{Q}}_1}, \cdots ,{{\bf{Q}}_K}] \in {\cal W}$. Using (\ref{overallmapping}) and Definition 1, the NE of the game can now be characterized by the following fixed-point equation:
\begin{equation}\label{fixedpoint}
    {\bf{F}}({{\bf{ Q}}^\star}) = {{\bf{ Q}}^\star}.
\end{equation}

According to Proposition 1.1 of \cite{Bertsekas-1989}, the contraction mappings have a unique fixed point. It is formally stated as the following lemma.

 \itshape \textbf{Lemma 2:}  \upshape  Game $ {\cal G}$  has a unique NE if the mapping ${\bf{F}}(\cdot)$  defined in (\ref{overallmapping}) satisfies:
\begin{equation}\label{lemma2}
    {\left\| {{\bf{F}}({{\bf{ Q}}^{(1)}}) - {\bf{F}}({{\bf{ Q}}^{(2)}})} \right\|_F} <  {\left\| {{{\bf{ Q}}^{(1)}} - {{\bf{ Q}}^{(2)}}} \right\|_F},
\end{equation}
for any two different  ${{\bf{ Q}}^{(1)}}$  and  ${{\bf{ Q}}^{(2)}}$ in the closed set ${\cal W}$.\hfill $\Box$

However, directly using this lemma is difficult. Instead, we give a sufficient condition for (\ref{lemma2}) to be satisfied, which  plays a key role in the study of uniqueness of the NE.

\itshape \textbf{Lemma 3:}  \upshape
The sufficient condition for (\ref{lemma2}) to hold is that, for any link $k$, the mapping defined in (\ref{mappingfunction}) satisfies
\begin{equation}\label{lemma2}
    {\left\| {{{\bf{F}}_k}({\bf{Q}}_{ - k}^{(1)}) - {{\bf{F}}_k}({\bf{ Q}}_{ - k}^{(2)})} \right\|_F} < \sqrt {\frac{1}{{K - 1}}} {\left\| {{\bf{ Q}}_{ - k}^{(1)} - {\bf{ Q}}_{ - k}^{(2)}} \right\|_F}, \forall  {\bf{Q}}_{ - k}^{(1)}, {\bf{Q}}_{ - k}^{(2)} \in {{\cal W}_{ - k}}.
\end{equation}

\itshape \textbf{Proof:} \upshape Our main task is to show that, under condition (\ref{lemma2}), the condition of the mapping ${\bf{F}}(\cdot)$ to be a contraction in (\ref{lemma2})  is satisfied. We have
\begin{align}
\left\| {{\bf{F}}({{\bf{Q}}^{(1)}}) - {\bf{F}}({{\bf{Q}}^{(2)}})} \right\|_F^2& = \sum\limits_{k = 1}^K \left\| {{{\bf{F}}_k}({\bf{Q}}_{ - k}^{(1)}) - {{\bf{F}}_k}({\bf{Q}}_{ - k}^{(2)})} \right\|_F^2 \\
 &< \frac{1}{{K - 1}} \sum\limits_{k = 1}^K \left\| {{\bf{Q}}_{ - k}^{(1)} - {\bf{Q}}_{ - k}^{(2)}} \right\|_F^2\label{secondin} \\
 &= \left\| {{{\bf{Q}}^{(1)}} - {{\bf{Q}}^{(2)}}} \right\|_F^2\label{lastinequ}
\end{align}
where (\ref{secondin}) follows from (\ref{lemma2}), and (\ref{lastinequ}) follows by  the definitions of ${{\bf{ Q}}_{ - k}}$  and $\bf{ Q}$.  \hfill $\Box$

Based on the above results, we give a sufficient condition to guarantee the uniqueness of the equilibrium for the case of full column-rank matrices in the following theorem:

\itshape \textbf{Theorem 2:}  \upshape Define  ${{\bf{T}}_k} \buildrel \Delta \over = {\left( {{\bf{I}} + {P_{\rm{T}}}\sum\limits_{i = 1}^K {{{\bf{H}}_{i,k}}{\bf{H}}_{i,k}^H} } \right)^{ - 1}}$. Suppose that the channel matrices ${\left\{ {{{\bf{H}}_{k,k}}} \right\}_{k \in {\Psi _K}}}$ are full column rank, i.e., ${\rm{rank}}({{\bf{H}}_{k,k}}) = M$. Let ${\alpha _k}$ be
\begin{equation}\label{defineafak}
    {\alpha _k}  = \frac{{\rho ({\bf{H}}_{k,k}^H{{\bf{H}}_{k,k}}){{\left\| {{{\bf{D}}_{{{\bf{Q}}_{ - k}}}}{{\bf{R}}_k}} \right\|}_2}}}{{{{\left( {{\lambda _{\min }}\left( {{\bf{H}}_{k,k}^H{{\bf{T}}_k}{{\bf{H}}_{k,k}}} \right)} \right)}^2}}},
\end{equation}
where ${{\bf{D}}_{{{\bf{Q}}_{ - k}}}}{{\bf{R}}_k}$  is the Jacobian matrix of ${{\bf{R}}_k}$  w.r.t. ${{\bf{Q}}_{ - k}}$  and is given by
\begin{equation}\label{Jociab}
     {{\bf{D}}_{{{\bf{Q}}_{ - k}}}}{\bf{R}}_k = \left[ {{\bf{H}}_{1,k}^* \otimes {{\bf{H}}_{1,k}}, \cdots ,{\bf{H}}_{k - 1,k}^* \otimes {{\bf{H}}_{k - 1,k}},{\bf{H}}_{k + 1,k}^* \otimes {{\bf{H}}_{k + 1,k}}, \cdots ,{\bf{H}}_{K,k}^* \otimes {{\bf{H}}_{K,k}}} \right]
\end{equation}
Then for sufficient large maximum transmit power $P_{\rm{T}}$, the NE of game ${\cal G}$  is unique if
\begin{equation}\label{uniquecondition}
{\alpha _k}< \sqrt{\frac{1}{{K - 1}}},\forall k \in {\Psi _K}.
\end{equation}

\itshape \textbf{Proof:}  \upshape  Please see Appendix \ref{proofTheorem2}.   \hfill $\Box$

The importance of Condition (\ref{uniquecondition}) is that for given direct channel gains, it explicitly qualifies how small the multi-link interference  each link can tolerate to guarantee the uniqueness of the NE. Hence, Condition (\ref{uniquecondition}) can be checked in practice to facilitate the admission control.  In contrast, \cite{Miao-2011} only showed that the number of the equilibria is determined by the cross-channel gains and the direct channel gains, without quantifying how they are related with each other.

2) \emph{Case of more general channel matrices}

In practical systems, the channel matrices may not be full column rank.  In this part, we consider the more general case without making any restrictive assumptions on the channel structure.

For each link $k$, we write the eigendecomposition of ${{\bf{H}}_{k,k}}$ as ${{\bf{H}}_{k,k}}{\rm{ = }}{{{\bf{\bar U}}}_k}{\bf{{\bar \Lambda }}_k}{\bf{\bar V}}_k^{\rm{H}}$, where ${{{\bf{\bar U}}}_k} \in {\mathbb{C}^{N \times {r_k}}},{{{\bf{\bar V}}}_k} \in {\mathbb{C}^{M \times {\bar r_k}}}$ are semi-unitary matrices, ${\bf{{\bar \Lambda }}_k} \in {\mathbb{C}^{{\bar r_k} \times {\bar r_k}}}$ is a diagonal matrix  with positive eigenvalues, and $\bar r_k$ is the rank of matrix  ${{\bf{H}}_{k,k}}$, i.e., ${\bar r_k} = {\rm{rank}}({{\bf{H}}_{k,k}}) \le {\rm{min}}\{ M,N\}$. To maximize each link's  EE in (\ref{EE}),   each link $k$'s optimal covariance matrix  should lie in the subspace orthogonal to the null-space of ${{\bf{H}}_{k,k}}$ for a given ${{\bf{Q}}_{ - k}}$. It follows that the best response of the transmit covariance matrix of each link $k$ belongs to the following class of matrices:
\begin{equation}\label{optimalset}
{{\bf{Q}}_k} = {{{\bf{\bar V}}}_k}{{{\bf{\bar Q}}}_k}{\bf{\bar V}}_k^{\rm{H}},\forall k \in {\Psi _K},
\end{equation}
with
\begin{equation}\label{set}
{{{\bf{\bar Q}}}_k} \in {{\bar {\cal W}}_k} \buildrel \Delta \over = \left\{ {{\bf{X}} \in {\mathbb{C}^{{r_k} \times {r_k}}}|{\bf{X}} \in \mathbb{S}_ + ^{r_k \times r_k},{\rm{tr}}\left( {\bf{X}} \right) \le {P_{\rm{T}}}} \right\}.
\end{equation}
By inserting (\ref{optimalset}) into game ${\cal G}$ in (\ref{game}) and defining ${{{\bf{\bar H}}}_{j,k}} = {{\bf{H}}_{j,k}}{{{\bf{\bar V}}}_j},\forall j,k$, game ${\cal G}$ can be transformed into the following lower-dimensional game ${\bar {\cal G}}$, given as
\begin{equation}\label{simplegame}
    \begin{array}{*{20}{c}}
{({\bar {\cal G}}):}&\begin{array}{l}
\mathop {\max }\limits_{{{\bf{\bar Q}}_k}} \ {\rm{E}}{{\rm{E}}_k}\left( {{{{\bf{\bar Q}}}_k},{{{\bf{\bar Q}}}_{ - k}}} \right) = \frac{{{{\log }_2}\left| {{\bf{I}} + {\bf{\bar H}}_{k,k}^H{\bf{\bar R}}_k^{ - 1}{{{\bf{\bar H}}}_{k,k}}{{{\bf{\bar Q}}}_k}} \right|}}{{{\rm{tr}}({{{\bf{\bar Q}}}_k}) + {P_{\rm{C}}}}}\\
{\rm{s}}{\rm{.t}}{\rm{. }}\quad {{\bf{\bar Q}}_k} \in {{\bar {\cal W}}_k},
\end{array}&{\forall k \in {\Psi _K}},
\end{array}
\end{equation}
where ${{{\bf{\bar Q}}}_{ - k}} = {{{\bf{\bar Q}}}_1} \times  \cdots  \times {{{\bf{\bar Q}}}_{k - 1}} \times {{{\bf{\bar Q}}}_{k + 1}} \times  \cdots  \times {{{\bf{\bar Q}}}_K}$ and ${\bf{\bar R}}_k^{ - 1} = {\bf{\bar R}}_k^{ - 1}({{{\bf{\bar Q}}}_{ - k}}) = {\bf{I}} + \sum\nolimits_{j \ne k} {{{{\bf{\bar H}}}_{j,k}}} {{{\bf{\bar Q}}}_j}{\bf{\bar H}}_{j,k}^H$.
Now the channel matrices ${{{\bf{\bar H}}}_{k,k}} \in {\mathbb{C}^{N \times {r_k}}}, \forall k$ are full column rank. Hence, we can apply the same derivations for Theorem 2 to obtain the uniqueness condition for this more general channel case, which is the same as that in (\ref{uniquecondition}) except that the channel matrices ${{\bf{H}}_{j,k}}\forall j,k$ are replaced by  ${{{\bf{\bar H}}}_{j,k}}\forall j,k$.

\begin{figure}[h]
\centering
\includegraphics[width=4.5in]{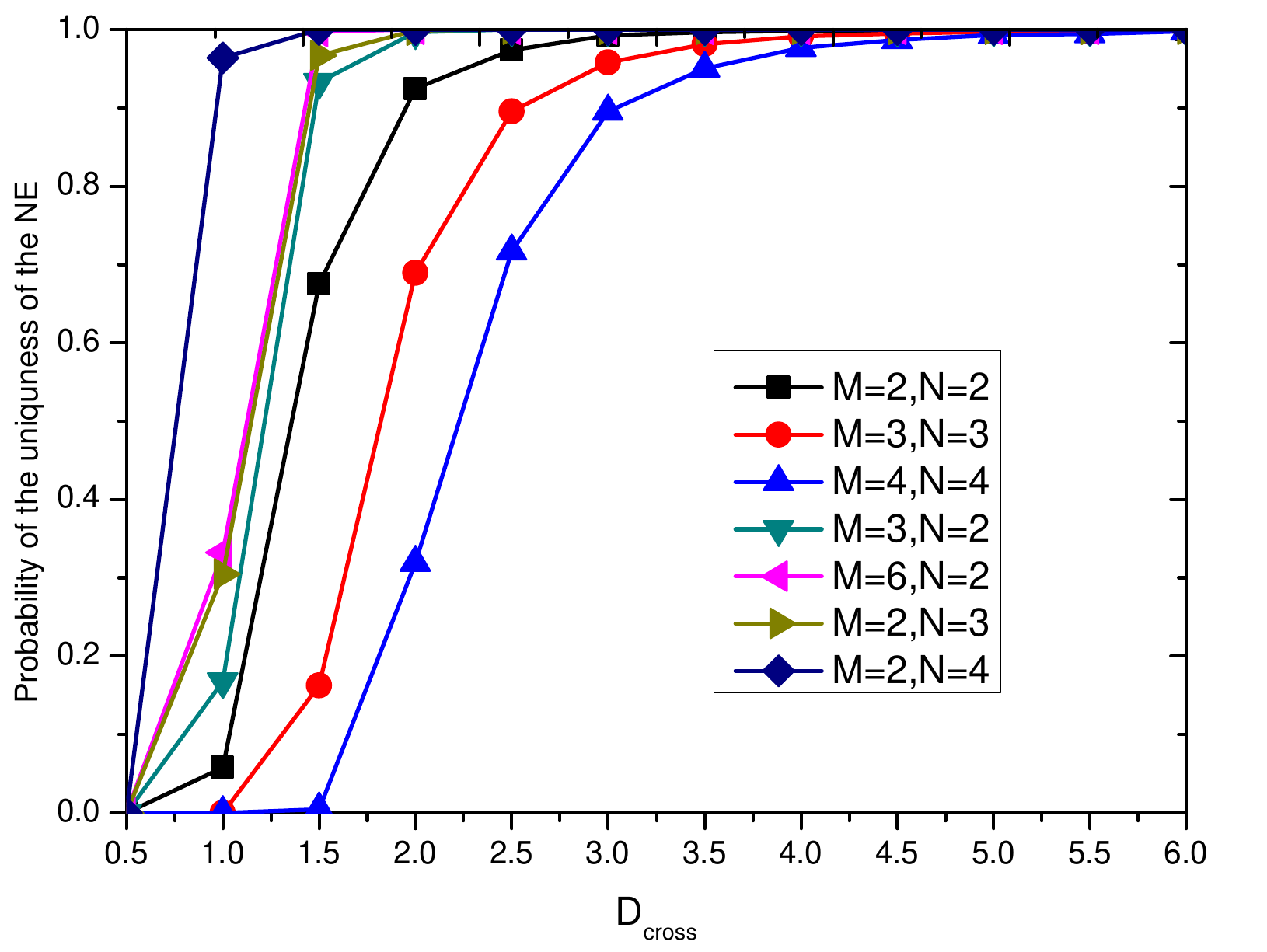}   \caption{Probability of the uniqueness of the NE for different numbers of antennas.}
\label{Unipro}
\end{figure}

To give more insights into the physical interpretation of the uniqueness conditions of the NE,  Fig.~\ref{Unipro} is plotted to show the probability for the uniqueness conditions to be satisfied for different cases. Specifically, full-rank square, fat and tall channel matrices are simulated. If the channel matrices are square or tall, condition (\ref{uniquecondition}) in Theorem 2 is applied to check the uniqueness condition of the NE. In the case of fat channel matrices, the above derived condition for general channel matrices is used to check the uniqueness condition of the NE. The probability of the uniqueness of the NE is defined as the ratio of the number of channel matrices that guarantee the uniqueness condition to the total number of channel matrices. For simplicity, we consider a symmetric system with two links ($K=2$) where the direct-channel distances for both links are set to be $D_{\rm{direct}}=1$ and both links have the same cross-channel distances (i.e., $D_{1,2}=D_{2,1}=D_{\rm{cross}}$). The power constraint and the noise power are set to be $P_T=10^{-2}{\rm{W}}$, $\sigma^2=10^{-3}{\rm{W}}$, respectively. Note that $P_T\gg \sigma^2$. These results are obtained by testing over 10000 random channel matrices whose entries are generated as the circularly symmetric, zero-mean, complex Gaussian random variables with variance equal to the square root of the channel path loss power, where the path loss exponent is assumed to be 3.5. Several interesting observations can be found from Fig.~\ref{Unipro}: 1) The uniqueness probability of the NE for all cases increase with the cross-channel distance corresponding the decrease in the interference. This is reasonable since in the extreme case when the cross-channel distance approaches infinity, the system becomes two independent point-to-point links. On the other hand, when one of the receivers is too close to one of its unintended transmitters, one of the two links should be shut down, which is instructive for practical use;  2) For the case of square channel matrices, the uniqueness probability decreases with the number of antennas; 3) For the case of fat channel matrices, increasing the number of transmitter antennas while keeping the number of receiver antennas fixed leads to an increase in the uniqueness probability of the NE. For example, the curve associated to the case of $2\times5$ MIMO channels is higher than that associated to the case of $2\times3$ MIMO channels for any given $D_{\rm{cross}}$. Similar observations hold for the case of tall channel matrices. This observation is of significant importance: equipping unequal number of antennas at transmitters and receivers can dramatically improve the uniqueness condition of the NE.

\section{Asynchronous Distributed Energy Efficient Algorithm}\label{algorithm}

To reach the NE of game ${\cal G}$, we employ the totally asynchronous algorithm \cite{Bertsekas-1989}. The main characteristic of the asynchronous algorithm is that some users are allowed to update their best response more frequently than others. This algorithm has been successfully employed to deal with the rate maximization game in \cite{scutari2008competitive,Scutari-2009}.

For the sake of readability, we briefly introduce the asynchronous algorithm and adapt it to our EE maximization problem. To this end, we first introduce some definitions and assume that the set of times at which the links update their solutions is  the discrete set ${\cal T}  = \{ 0,1,2, \cdots \} $. Let ${\cal T} {^k} \subseteq {\cal T}$  be the subset of times at which transmitter $k$  updates its solution and let ${{\bf{Q}}_k}(t)$ be the updated transmit covariance matrix of transmitter $k$ at time $t \in {\cal T}{^{{\rm{ }}k}}$. Denote $\tau _r^k(t)$ as the most recent time at which the interference from transmitter $r$ is measured at receiver $k$ at time $t$. Hence, at time $t$, link $k$ updates its transmit covariance matrix based on the interference from
\begin{equation}\label{inter}
  {{\bf{Q}}_{ - k}}({\tau ^k}(t)) \buildrel \Delta \over = \left( {{{\bf{Q}}_1}(\tau _1^k(t)), \cdots ,{{\bf{Q}}_{k - 1}}(\tau _{k - 1}^k(t)),{{\bf{Q}}_{k + 1}}(\tau _{k + 1}^k(t)), \cdots ,{{\bf{Q}}_K}(\tau _K^k(t))} \right).
\end{equation}

To guarantee that the system is totally asynchronous, three conditions should be satisfied by the  schedules $\{ \tau _r^q(t)\} $ and $\{ {\cal T}{^{{\rm{ }}k}}\} $ \cite{Bertsekas-1989}: A1) $0\leq\tau _r^k(t)\leq t$; A2) ${\lim _{k \to \infty }}\tau _r^k({t_k}) =  + \infty $; and A3) $\left| {{{\cal T}^k}} \right| = \infty $; where $\left\{ {{t_k}} \right\}$ is a sequence of ${\cal T} {^k}$.
Based on the above definitions, the asynchronous distributed EE algorithm is described in Algorithm 1, where ${T_{\max }}$ is the maximum number of the iterations.
\begin{algorithm}
\caption{\itshape Asynchronous Distributed Energy-Efficiency (ADEE) Algorithm}
\begin{algorithmic}[1]
\STATE Set $t = 0$  and initialize ${{\bf{Q}}_k}(0) \in {{\cal W}_k},\forall k$;
\STATE FOR $t = 0:{T_{\max }}$\\
       \quad FOR $k = 1:K$\\
       \quad \quad IF $t \in {{\cal T}^k}$\\
       \quad \quad \quad Update  ${{\bf{Q}}_k}(t + 1)$  using
        Algorithm 2 in the following subsection based on ${{\bf{Q}}_{ - k}}({\tau ^k}(t))$.\\
       \quad \quad ELSE\\
       \quad \quad \quad ${{\bf{Q}}_k}(t + 1) = {{\bf{Q}}_k}(t)$,\\
       \quad \quad END\\
       \quad END\\
       END\\
\end{algorithmic}
\end{algorithm}

\itshape \textbf{Remark 1 - Two special cases:}  \upshape Note that some well-known algorithms such as sequential and simultaneous algorithms \cite{Bertsekas-1989}, where the links update their solutions sequentially and simultaneously, are special cases of our algorithm. Each one of them can be obtained by appropriately choosing the scheduling  parameters $\{ {{\cal T}^k}\} $ and $\{ \tau _r^q(t)\}$. In the sequential  algorithm, all links update their own strategies sequentially,  whose scheduling parameters  $\{ {{\cal T}^k}\} $ and $\{ \tau _r^q(t)\}$ are chosen as
${\cal T}{^{{\rm{ }}k}} = \{ mK + k,{\rm{ }}m \in {{\cal N}_ + }\} {\rm{ = }}\{ k,K + k,2K + k, \cdots \} ,\tau _r^k(t) = t,\forall k,r$, where ${{\cal N}_ + }$ denotes the discrete set ${{\cal N}_ + } = \{ 0,1,2, \cdots \} $. Similarly, in the simultaneous algorithm, all links update their own strategies simultaneously  with the scheduling parameters chosen as
${\cal T}{^{{\rm{ }}k}} = {{\cal N}_ + },\tau _r^k(t) = t,\forall k,r$. In contrast, for the DBFEE \cite{ChenziJiang-2013} and TLEE algorithms \cite{ShiwenTSP}, each transmitter should collect the knowledge of the channels to all the receivers over the network at the initial stage of the algorithm. Moreover, in each iteration all the receivers should feed back the complex-valued matrices to all the transmitters in the network, which is not scalable for large scale networks.

\subsection{Per-link Response Problem}

The best response problem for link  $k$ can be written as \footnote{As ${{\bf{Q}}_{ - k}}$  is treated as constant, ${{C_k}({{\bf{Q}}_k})}$ is used to represent ${{C_k}({{\bf{Q}}_k},{{\bf{Q}}_{ - k}})}$  for simplicity.}
\begin{equation}\label{perlinkproblem}
\begin{array}{l}
  \mathop {\max }\limits_{{{\bf{Q}}_k}\in {{\cal W}_k}} \ {\rm{ }}\frac{{{C_k}({{\bf{Q}}_k})}}{{{\rm{tr(}}{{\bf{Q}}_k}{\rm{)}} + {P_C}}}.
\end{array}
\end{equation}
Since the numerator and denominator in the objective function of Problem (\ref{perlinkproblem}) are concave and affine in ${\bf{Q}}_k$  respectively, the objective function of (\ref{perlinkproblem}) is a pseudo-concave function \cite{Zappone-2014}. Moreover, the constraint in (\ref{perlinkproblem}) is convex. Hence, problem (\ref{perlinkproblem}) can be solved by the following lemma, the proof of which can be found in Proposition 6 of \cite{Schaible-1983}.

\itshape \textbf{Lemma 4:}  \upshape Define function $G(\kappa )$  as
\begin{equation}\label{fkafa}
G(\kappa ) \buildrel \Delta \over = \mathop {\max }\limits_{{{\bf{Q}}_k} \in {{\cal W}_k}} \log_2 |{\bf{I}} + {\bf{H}}_{k,k}^H{\bf{R}}_k^{ - 1}{{\bf{H}}_{k,k}}{{\bf{Q}}_k}|
 - \kappa ({\rm{tr(}}{{\bf{Q}}_k}{\rm{) + }}{P_{\rm{C}}}),
\end{equation}
For fixed  $\kappa $, the solution of (\ref{fkafa}) is denoted as  ${\bf{Q}}_k^\star(\kappa )$. Then, solving (\ref{perlinkproblem}) is equivalent to finding the root of the equation
$G({\kappa ^\star}) \buildrel \Delta \over = {C_k}({\bf{Q}}_k^\star({\kappa ^\star})) - {\kappa ^\star}({\rm{tr(}}{\bf{Q}}_k^\star({\kappa ^\star}){\rm{) + }}{P_{\rm{C}}}){\rm{ = }}0$.\hfill $\Box$


Lemma 4 gives us insights to solve (\ref{perlinkproblem}). We can solve Problem (\ref{fkafa}) with fixed $\kappa $ firstly, while the optimal $\kappa $ can be searched via the Dinkelbach method \cite{Dinkelbach-1967}.

Now we attempt to solve (\ref{fkafa}) for fixed $\kappa$. To this end, we first write the  eigenvalue decomposition (EVD) of ${\bf{H}}_{k,k}^H{\bf{R}}_k^{ - 1}{{\bf{H}}_{k,k}}$  for each $k \in {\Psi _K}$  as
\begin{equation}\label{eachas}
  {\bf{H}}_{k,k}^H{\bf{R}}_k^{ - 1}{{\bf{H}}_{k,k}} = {{\bf{U}}_k}{{\bf{D}}_k}{\bf{U}}_k^{\rm{H}},
\end{equation}
where  ${{\bf{U}}_k}$ is a semi-unitary matrix of the eigenvectors with ${r_k} \buildrel \Delta \over = {\rm{rank(}}{\bf{H}}_{k,k}^H{\bf{R}}_k^{ - 1}{{\bf{H}}_{k,k}}{\rm{)}}$, ${{\bf{D}}_k} \in\mathbb{R} _{ +  + }^{r_k \times r_k}$ is a diagonal matrix with  ${{\bf{D}}_k} = {\rm{diag}}\{ {d_{k,1}}, \cdots ,{d_{k,r_k}}\}$ being the eigenvalues.

Then, given $k \in {\Psi _K}$ and ${{\bf{Q}}_{ - k}} \in {{\cal W}_{ - k}}$, the solution to problem (\ref{fkafa})  with fixed $\kappa$ is  \cite{kim2011optimal}:
 \begin{equation}\label{waterfilling}
{{\bf{Q}}_k} = {{\bf{U}}_k}{\bf{\Lambda}}_k {\bf{U}}_k^H,
 \end{equation}
where ${\bf{\Lambda}}_k= {\rm{diag}}\{ {q_{k,1}}, \cdots ,{q_{k,{r_k}}}\}$ represents the power allocations on all subchannels with
 \begin{equation}\label{isgivenby}
    {q_{k,m}} = {\left[ {\frac{1}{{(\kappa  + {\lambda _k}) \ln 2}} - \frac{1}{{{d_{k,m}}}}} \right]^ + },m \in \{ 1,2, \cdots ,{r_k}\},
\end{equation}
where ${\lambda _k} \ge 0$ is the Lagrange multiplier associated with the power constraint, which should be chosen to satisfy the complementarity slackness condition:  ${\lambda _k}\left( {{\rm{tr(}}{{\bf{Q}}_k}{\rm{)}} - {P_{\rm{T}}}} \right) = 0$.

After solving problem (\ref{fkafa}), we utilize the Dinkelbach method \cite{Dinkelbach-1967} to update $\kappa $ as follows
\begin{equation}\label{updateka}
{\kappa _{n + 1}} = \frac{{{C_k}({\bf{Q}}_k^{(n)\star}({\kappa _n}))}}{{{\rm{tr(}}{\bf{Q}}_k^{(n)\star}({\kappa _n}){\rm{) + }}{P_{\rm{C}}}}},n=1,2,\cdots,
\end{equation}
where $n$ is the iteration index.

To summarize the above analysis, we give the following algorithm to solve the per-link problem in (\ref{perlinkproblem}).
\begin{algorithm}[H]
\caption{\itshape The Dinkelbach method to solve (\ref{perlinkproblem})}
\begin{algorithmic}[1]
\STATE Initialization: ${\kappa _0}$ satisfying $G({\kappa _0}) \ge 0$, tolerance $\varepsilon $, iteration number $n = 0$;
\STATE For given ${\kappa _n}$, solve  (\ref{fkafa}) to get the optimal ${\bf{Q}}_k^{(n)\star}({\kappa _n})$;
\STATE If $\left| {G({\kappa _n})} \right| > \varepsilon $, update ${\kappa _{n + 1}}$  in (\ref{updateka}), $n = n + 1$, go back to step 2. Otherwise, terminate.

\end{algorithmic}
\end{algorithm}

\itshape \textbf{Remark 2 - Distributed nature of the algorithm:}  \upshape For transmitter $k$ to update ${\bf{Q}}_k$, it requires receiver $k$ to feed back the channel matrix ${{\bf{H}}_{k,k}}$ and the IPN covariance matrix  ${{\bf{R}}_k}$ according to (\ref{fkafa}). It is well-known that the channel matrix ${{\bf{H}}_{k,k}}$ can be estimated at the receiver. For the IPN covariance matrix  ${{\bf{R}}_k}$, it can be easily computed at the receiver as follows: First, transmitter $k$ sends a sequence of training sequence to receiver $k$. The statistical information of the training sequence is assumed to be known at the receiver. Then the IPN covariance matrix can be obtained by subtracting the covariance matrix of the received training sequence from the covariance matrix of the total received signal \footnote{Note that this has been done in most of the existing distributed algorithms \cite{kim2011optimal,Nguyen2014MAC,Nguyen2014BC,Yanqing-2014,ChenziJiang-2013,Sigen-2003,scutari2008competitive,Scutari-2009,scutari2010mimo,wang2011robust,Nguyen2011TSP,Nguyen2014BD,zhou2014network}.}.
 Hence, the ADEE algorithm can be performed in a totally distributed and asynchronous way without the need of information exchange among different links. In contrast, for the DBFEE \cite{ChenziJiang-2013} and TLEE algorithms \cite{ShiwenTSP},  in each iteration all the receivers should feed back the complex-valued matrices to all the transmitters in the network, which is not scalable for large scale networks.

\subsection{Convergence Analysis}

In this section, the sufficient condition that guarantees the global convergence of the ADEE algorithm is given. Interestingly, we find that this convergence condition is the same as the uniqueness condition obtained in Theorem 2, as proved in the following.

\itshape \textbf{Theorem 3:}  \upshape In the case of full column-rank channel matrices, suppose that condition (\ref{uniquecondition}) in Theorem 2 is satisfied and the maximum transmit power is large enough, then as ${T_{\max }} \to \infty $, the sequence generated by the ADEE algorithm converges to the unique NE of game ${\cal G}$, for any given set of feasible initial conditions and updating schedules.

\itshape \textbf{Proof:}  \upshape Please see Appendix \ref{proofTheorem3}.    \hfill $\Box$

The proof for the more general case can be derived similarly. It is omitted for brevity.

\itshape \textbf{Remark 3 - Robustness of the algorithm:}  \upshape The condition for the convergence of the ADEE algorithm is independent of the update schedule for each link. Hence, all special cases of the ADEE algorithm like the well-known sequential and simultaneous updating are guaranteed to converge under the same condition in (\ref{uniquecondition}). Furthermore, there is no restrict constraints on the updating schedule on each link, so that some links are allowed to update their transmit covariance matrices more often than others, without affecting the convergence of the algorithm.

\section{Performance Analysis of the Proposed Algorithm}\label{performanceana}
In this section, we give the performance analysis for the proposed algorithm. Firstly, the impact of the circuit power consumption on the system performance is studied. Then, we investigate the  tradeoff between SE and EE.

\subsection{Impact of Circuit Power Consumption on the System Performance}\label{circuitimpact}

In this part, we investigate the impact of circuit power consumption on the overall SE and EE performance. Due to the coupling interference among different links, it is difficult to provide the analytical results. To simplify the analysis, we only consider the extreme case that  all the links are separated far away with each other so that the interference among all the links reduces to almost zero. In this case, the overall system can be regarded as $K$ independent links. Then, we only need to study the impact of circuit power consumption on the each link $k$'s SE and EE performance, which is given in the following theorem.

\itshape \textbf{Theorem 4:}  \upshape When the links are separated far away and no power constraints are imposed at the transmitters, the maximum achievable EE of each link $k$ for a given circuit power consumption  $P_C$ decreases with $P_C$, but the corresponding SE of each link $k$ increases with $P_C$.

\itshape \textbf{Proof:}  \upshape Please see Appendix \ref{proofTheorem4}.   \hfill $\Box$

\begin{figure}[h]
\centering
\includegraphics[width=4.5in]{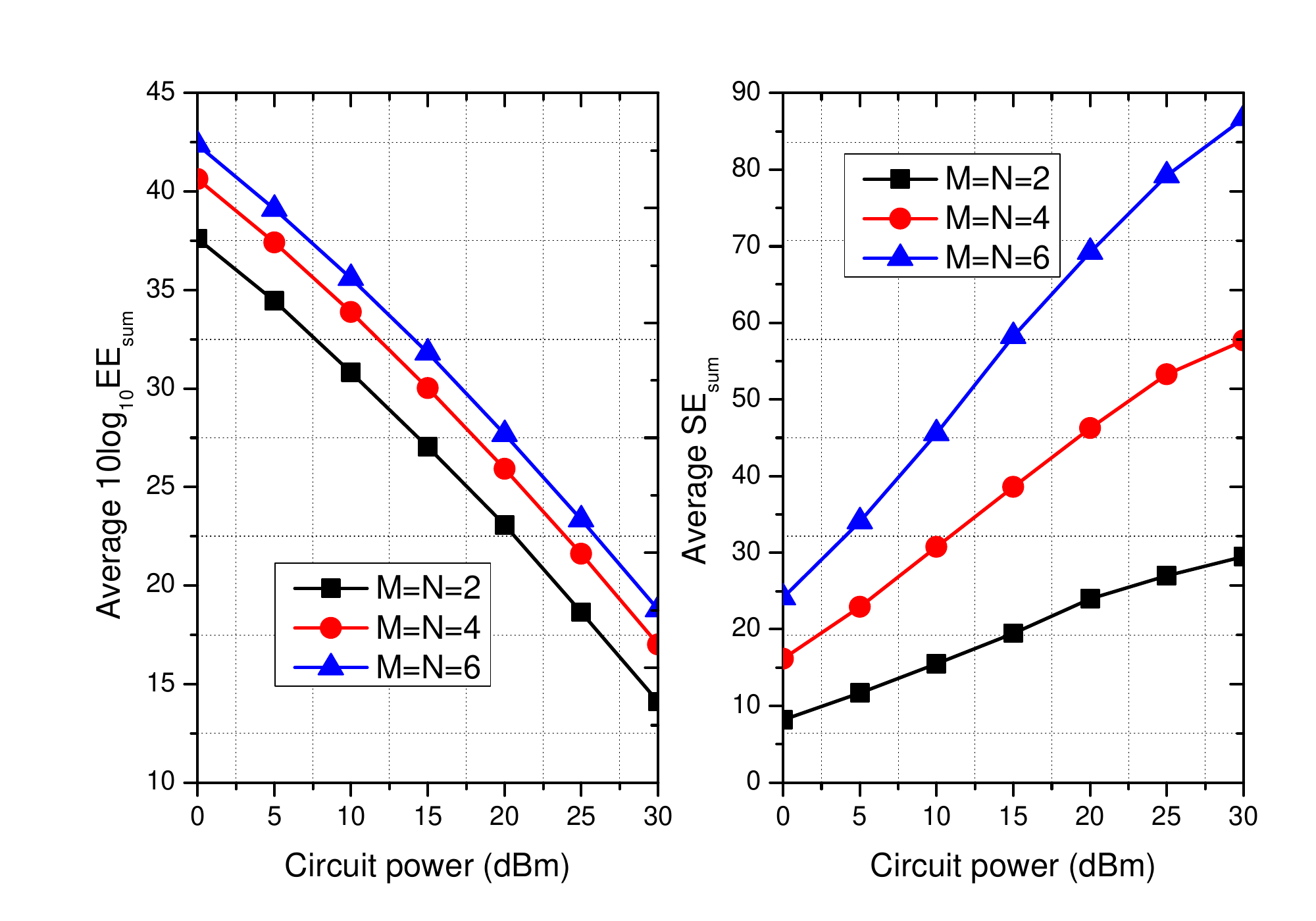}   \caption{The impact of circuit power consumption on system's SE and EE when links are separated far away.}\vspace{-0.8cm}
\label{vali}
\end{figure}

To validate the analysis in Theorem 4, we plot a figure in Fig.~\ref{vali} to show the impact of circuit power consumption on system's SE and EE for the same scenario in Fig.~\ref{Unipro} with $D_{\rm{cross}}=5$. The sum-EE and the sum-SE are defined as ${\rm{E}}{{\rm{E}}_{{\rm{sum}}}} = \sum\nolimits_{k = 1}^K {{\rm{E}}{{\rm{E}}_k}} $, ${\rm{S}}{{\rm{E}}_{{\rm{sum}}}} = \sum\nolimits_{k = 1}^K {{C_k}} $, respectively. For illustration purpose, the y-axis is shown by $10{\rm{log}}_{10}{\rm{EE}}_{\rm{sum}}$.  It can be seen from the figure that the sum-EE decreases with the circuit power consumption while the corresponding sum-SE increases with it for all considered numbers of antennas, which validates the correction of the theorem. From the simulation section, the above trend also holds for the general case with sizeable interference among the links, though the above theorem is derived for the extreme case.

\subsection{Tradeoff between Spectral and Energy Efficiency}\label{tradeoff}

In this part, we study the tradeoff between SE and EE for the proposed ADEE algorithm and the asynchronous distributed spectral efficient (ADSE) algorithm \cite{Scutari-2009}. Generally, it is difficult to study this problem analytically due to the coupling interference. For this reason, we investigate the tradeoff for two special cases: low  transmit power constraint and  high transmit power constraint.

\subsubsection{Low transmit power constraint: $P_{\rm{T}}\rightarrow 0 $} In this case, the interlink interference can be neglected, i.e., ${{\bf{R}}_k} \approx {\bf{I}}, \forall k$, and the network reduces to $K$ independent point-to-point links. Hence, we only need to analyze one link's performance. Without loss of generality, we only focus on the performance of link $k$.
Denote the transmit power of link $k$ of the ADEE algorithm as $P_k \in [0,{P_{\rm{T}}}]$. Since $P_{\rm{T}}\rightarrow 0 $, from (\ref{rewitecfunc}) there is only one stream for link $k$ that transmits with positive power. Hence, the SE for link $k$ can be written as ${C_k}\left( {P_k} \right) = {\log _2}\left( {1 + {d_{k,1}}P_k} \right)$, where $d_{k,1}$ is the maximum eigenvalue of ${\bf{H}}_{k,k}^H{{\bf{H}}_{k,k}}$. The optimal EE for link $k$ can be written as
\begin{equation}\label{optimalEEk}
 {\rm{E}}{{\rm{E}}_k}\left( {P_k} \right) = \frac{{{C_k}\left( {P_k} \right)}}{{P_k + {P_{\rm{C}}}}}.
\end{equation}
Then, we have
\begin{equation}\label{firstderivate}
\mathop {{\rm{lim}}}\limits_{P_k \to 0} {\rm{E}}{{\rm{E'}}_k}\left( {P_k} \right) = \frac{{{d_{k,1}}}}{{\ln 2{P_{\rm{C}}}}} > 0,
\end{equation}
where ${\rm{E}}{{\rm{E'}}_k}\left( {P_k} \right)$ is the derivative of ${\rm{E}}{{\rm{E}}_k}\left( {P_k} \right)$ w.r.t. $P_k$. Hence, ${\rm{E}}{{\rm{E}}_k}\left( {P_k} \right)$ is increasing for $P_k \in [0,{P_{\rm{T}}}]$. Then, the optimal transmit power is $P_k^*=P_{\rm{T}}$, i.e., link $k$ transmits with its maximum power.

For the ADSE algorithm, each link maximizes its SE selfishly by always using its maximum power, i.e., $P_k^* = {P_{\rm{T}}}$. Hence, both the ADEE algorithm and ADSE algorithm achieve the same SE and EE.

\subsubsection{High transmit power constraint: $P_{\rm{T}}\rightarrow \infty $}

In this case, the interlink interference cannot be neglected any more. Due to the interference, the analysis becomes more difficult. To simplify the analysis and get insights, we consider a symmetric system similar to one in \cite{Miao-2011} for the SISO frequency selective interference channel. We assume that ${\bf{H}} = {{\bf{H}}_{k,k}},\forall k$ and ${{\bf{H}}_{j,k}} = \alpha {\bf{H}},\forall j \ne k$, where $\alpha $ is a constant. In this symmetric system, all links transmit with the same covariance matrices, i.e., ${\bf{Q}} = {{\bf{Q}}_k},\forall k$. Denote the transmit power as ${\rm{tr(}}{\bf{Q}}) = P$.

The overall network EE is
\begin{equation}\label{overalllEE}
{\rm{EE = }}\frac{{K{{\log }_2}\left| {{\bf{I}} + {{\bf{H}}^{\rm{H}}}{{\bf{R}}^{ - 1}}{\bf{HQ}}} \right|}}{{P + {P_{\rm{C}}}}},
\end{equation}
and the overall network SE is
\begin{equation}\label{overallSE}
 {\rm{SE = }}K{\log _2}\left| {{\bf{I}} + {{\bf{H}}^{\rm{H}}}{{\bf{R}}^{ - 1}}{\bf{HQ}}} \right|,
\end{equation}
where ${\bf{R}} = {\bf{I}} + (K - 1){\alpha ^2}{\bf{HQ}}{{\bf{H}}^{\rm{H}}}$.

For the ADSE algorithm, each link selfishly maximizes its own SE by using its maximum transmit power, i.e., ${\rm{tr(}}{\bf{Q}}) = P = {P_{\rm{T}}}$. When $P_{\rm{T}}\rightarrow \infty $, the interlink interference becomes very large. According to Theorem 3 of \cite{Sigen-2003}, one of the optimal solutions employs beamforming (1-stream signaling) for all links. Hence, the optimal covariance matrix can be written as ${\bf{Q}} = {P_{\rm{T}}}{\bf{w}}{{\bf{w}}^{\rm{H}}}$, where ${\bf{w}}$ is the beam direction with unit norm. Then, the network SE of the ADSE algorithm in the high transmit power is upper bounded by
\begin{equation}\label{SEadse}
{\rm{S}}{{\rm{E}}_{{\rm{ADSE}}}} = \mathop {\lim }\limits_{{P_{\rm{T}}} \to \infty } K{\log _2}\left( {1 + \frac{{g{P_{\rm{T}}}}}{{1 + {\alpha ^2}(K - 1)g{P_{\rm{T}}}}}} \right) = K{\log _2}\left( {1 + \frac{1}{{{\alpha ^2}(K - 1)}}} \right),
\end{equation}
where $g = {{\bf{w}}^{\rm{H}}}{{\bf{H}}^{\rm{H}}}{\bf{Hw}}$ is a constant. The corresponding EE is ${\rm{E}}{{\rm{E}}_{{\rm{ADSE}}}} = \mathop {\lim }\limits_{{P_{\rm{T}}} \to \infty } {{{\rm{S}}{{\rm{E}}_{{\rm{ADSE}}}}} \mathord{\left/
 {\vphantom {{{\rm{S}}{{\rm{E}}_{{\rm{ADSE}}}}} {({P_{\rm{T}}} + {P_{\rm{C}}})}}} \right.
 \kern-\nulldelimiterspace} {({P_{\rm{T}}} + {P_{\rm{C}}})}} = 0$, which  is not desirable from the EE point of view.

 On the other hand, for the ADEE algorithm,  the system will not use the maximum power to transmit due to the power value in the denominator of  (\ref{overalllEE}). Hence, the interlink interference may not be so large. Thus, the above derivations for the high interference scenario are not applicable and it is difficult to analyze the performance of the ADEE algorithm in this case. However, from the above discussion, we can conclude that its overall EE will increase when the transmit power constraint is low, and keeps constant when the transmit power constraint is high, which is significantly larger than that of the ADSE algorithm. The overall SE has the same trend as the overall EE.

\section{Simulation Results}\label{simulation}

We consider an ad hoc network contained in a $250{\rm{m}} \times 250{\rm{m}}$ square area, in which all links are randomly distributed.  The distance from one transmitter to its unintended receiver is at least  $35{\rm{m}}$. The channel is modeled by path-loss  \cite{Assumptions-2009} and independent Rayleigh fading with the complex normal distribution, ${\cal C}{\cal N}(0,1)$. The channel path-loss is modeled as  $38.46 + 35{\log _{10}}(d)$ \cite{Assumptions-2009}. Each channel realization is obtained by generating a random set of links' positions as well as fading channel realizations. It is assumed that the transmitters and the receivers have the same number of antennas. Unless otherwise specified, the other main system parameters are given in Table \ref{tab1}¡£ For comparisons, the metrics used are the sum-EE and the sum-SE, which are defined as ${\rm{E}}{{\rm{E}}_{{\rm{sum}}}} = \sum\nolimits_{k = 1}^K {{\rm{E}}{{\rm{E}}_k}} $, ${\rm{S}}{{\rm{E}}_{{\rm{sum}}}} = \sum\nolimits_{k = 1}^K {{C_k}} $, respectively.
 \begin{table}[H]
\renewcommand{\arraystretch}{1.1}
\caption{\label{parameter}Main simulation parameters}\vspace{-0.3cm}
\label{tab1}
\centering
\begin{tabular}{c|c}
\hline\hline
\textbf{Parameters}  & \textbf{Value} \\
\hline
Number of links  $K$  & 4 \\
\hline
Number of transmit antennas $M$   & 4 \\
\hline
Number of receiver antennas $N$   &  4 \\
\hline
Direct-channel distance ${D_{{\rm{direct}}}}$ & 80 {\rm{m}}\\
\hline
 Noise power ${\sigma ^2}$ & -106 dBm \\
\hline
Circuit power consumption  ${P_{{\rm{C}}}}$ & 23 dBm \cite{Arnold-2010}\\
\hline
Tolerance  $\varepsilon$ &  $10^{(-5)}$  \\
\hline
Maximum transmit power  ${P_{\rm{T}}}$ &   30 dBm  \\
\hline
 Maximum number of iterations  ${T_{\max }}$ & 20 \\
\hline
 Channel path loss model & $38.46 + 35{\log _{10}}(d)$ \cite{Assumptions-2009} \\
\hline\hline
\end{tabular}
\end{table}

\subsection{Convergence behavior of the ADEE algorithm}

\begin{figure}
\centering
\includegraphics[width=4.5in]{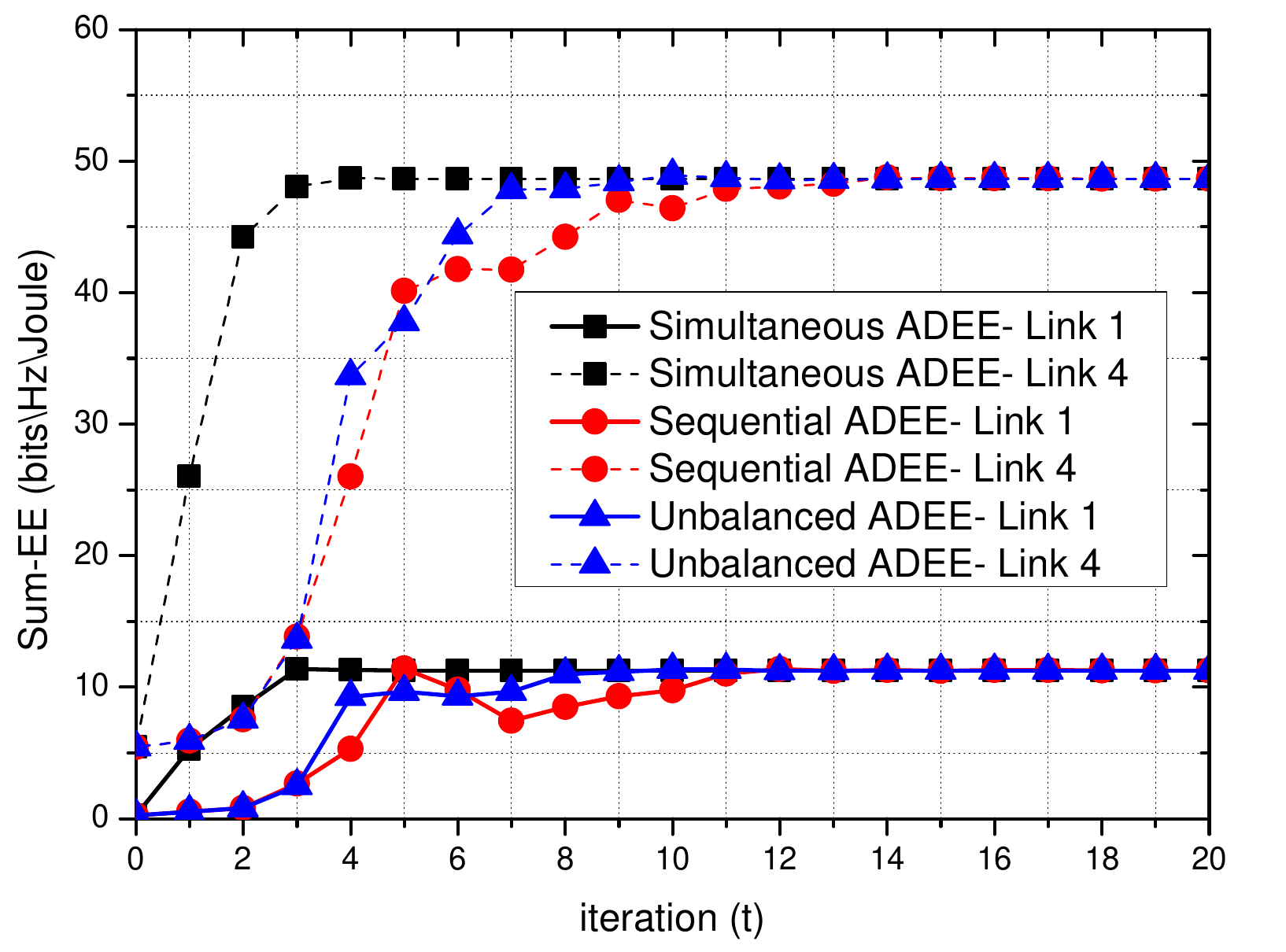}
\caption{Converegence behaviors for different updating schedules. }
\label{Convergence}
\end{figure}

\begin{figure}
\centering
\includegraphics[width=4.5in]{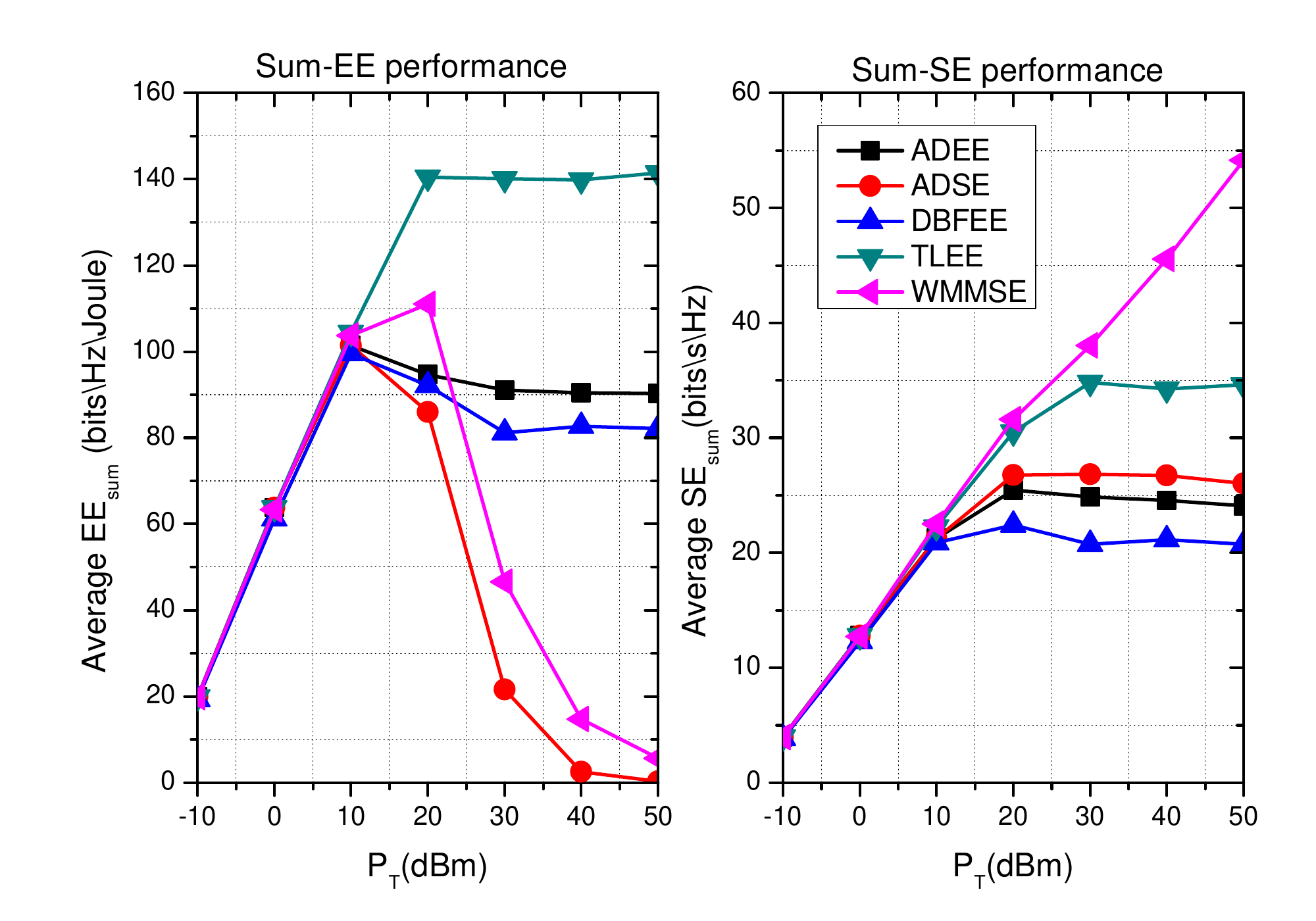}
\caption{Average sum-EE (left) and sum-SE (right) versus $P_T$ for various algorithms.}
\label{Sum-EESE-power}
\end{figure}

Fig.\ref{Convergence} illustrates the convergence behavior of the ADEE algorithm for different updating schedules for one randomly generated channel realization. For  comparison, the performance of another updating scheme, named `unbalanced ADEE', is also shown, where the parameters are chosen as
${{\cal T}^k} = \{ k,2k,3k, \cdots \} ,\tau _r^k(t) = t,\forall k,r$.
Hence, in the unbalanced ADEE, Link 1 is set to update its strategy faster than the other three links, and Link 4 is set to be the slowest to  update its strategy. Also, in each time, the number of links updating their strategies is no less than that of the sequential ADEE (only one link), no more than that of the simultaneous ADEE (all four links). To make the figure not too crowded, we report only the curves of two links (Link 1 and Link 4). It can be seen from this figure that the simultaneous ADEE converges faster than the other two schemes, and can converge within a few iterations. However, it takes about 12 iterations for the sequential ADEE to converge. The reason is that each user in the sequential ADEE is forced to wait for all the users scheduled in advance, before updating its own strategy. Moreover, the unbalanced ADEE converges a little faster than the sequential ADEE due to more links involved in updating their strategies in each iteration. From this figure, we find that different scheduling methods yield almost the same performance. For this reason, we only report the performance of the simultaneous ADEE due to its rapid convergence speed in the following simulations.

\subsection{Performance Comparison with Existing Algorithms}

We next compare the performance of the ADEE algorithm with some existing algorithms, including the ADSE algorithm \cite{Scutari-2009} where each link always uses its maximum transmit power, and DBFEE algorithm \cite{ChenziJiang-2013}. Since in the proposed ADEE algorithm each link only attempts to maximize its own EE selfishly, its achieved overall EE is generally suboptimal. Hence, it is interesting to study the performance gap between the proposed ADEE algorithm and the (near-)optimal sum-EE maximization algorithm. For this reason, we simulate the TLEE algorithm in \cite{ShiwenTSP}  that aims at the sum-EE maximization. Moreover, the WMMSE algorithm in \cite{Qingjiang-2011} that focuses on the sum-SE maximization is also simulated.

\subsubsection{Impact of the Maximum Transmit Power on the System Performance}

Fig.~\ref{Sum-EESE-power} shows the average sum-EE (left) and the average sum-SE (right) versus the maximum transmit power for various algorithms. It can be seen that in the low transmit power regime, i.e., $P_T\leq10$ dBm, both the sum-EE and the sum-SE of all algorithms monotonically increase with the maximum transmit power. Interestingly, we find that in this regime the proposed ADEE algorithm almost achieves the optimal sum-EE and sum-SE. That is, the performance gap with the optimal sum-EE and sum-SE performance approaches zero. Fortunately, many short-distance wireless networks operate in this regime, such as ad hoc networks \cite{Jianwei-2006,Biao-2006}, femtocells, and wireless sensor networks \cite{Akyildiz-2005}.  However, in the high transmit power regime, the sum-EE of the ADRM algorithm decreases dramatically, while its sum-SE performance becomes stable. This is because in the high transmit power regime, the system becomes interference limited and increasing the transmit power may slightly help increase the sum-SE, which leads to a significant performance loss in terms of the sum-EE due to the increased transmit power. Note that the sum-SE achieved by the ADEE algorithm is comparable with that achieved by the ADRM algorithm, which corroborates with the analysis in Section \ref{performanceana}. Fig.~\ref{Sum-EESE-power} also shows that the ADEE algorithm outperforms the DBF algorithm in terms of both the sum-EE and the sum-SE. This is because the DBFEE algorithm is primarily designed for the symmetry system and may not be suitable for the asymmetric ad hoc network considered here. Moreover, this algorithm  requires a substantial feedback overhead on the network and further the power consumption due to these information exchanges is not considered. As expected, both the sum-EE and the sum-SE performance of the TLEE algorithm is superior to that of the ADEE algorithm in the high transmit power regime due to the selfish nature of the ADEE algorithm. However, this benefit comes at the cost of the heavy information exchange overhead, high computational complexity, synchronization requirements of the networks, and the need for coordination among the users. In contrast, the proposed ADEE algorithm does not need information exchange among different links, and can be implemented in a totally distributed and asynchronous manner, which is appealing for some practical applications.

\subsubsection{Impact of the Number of Antennas on the System Performance}

\begin{figure}
\centering
\includegraphics[width=4.5in]{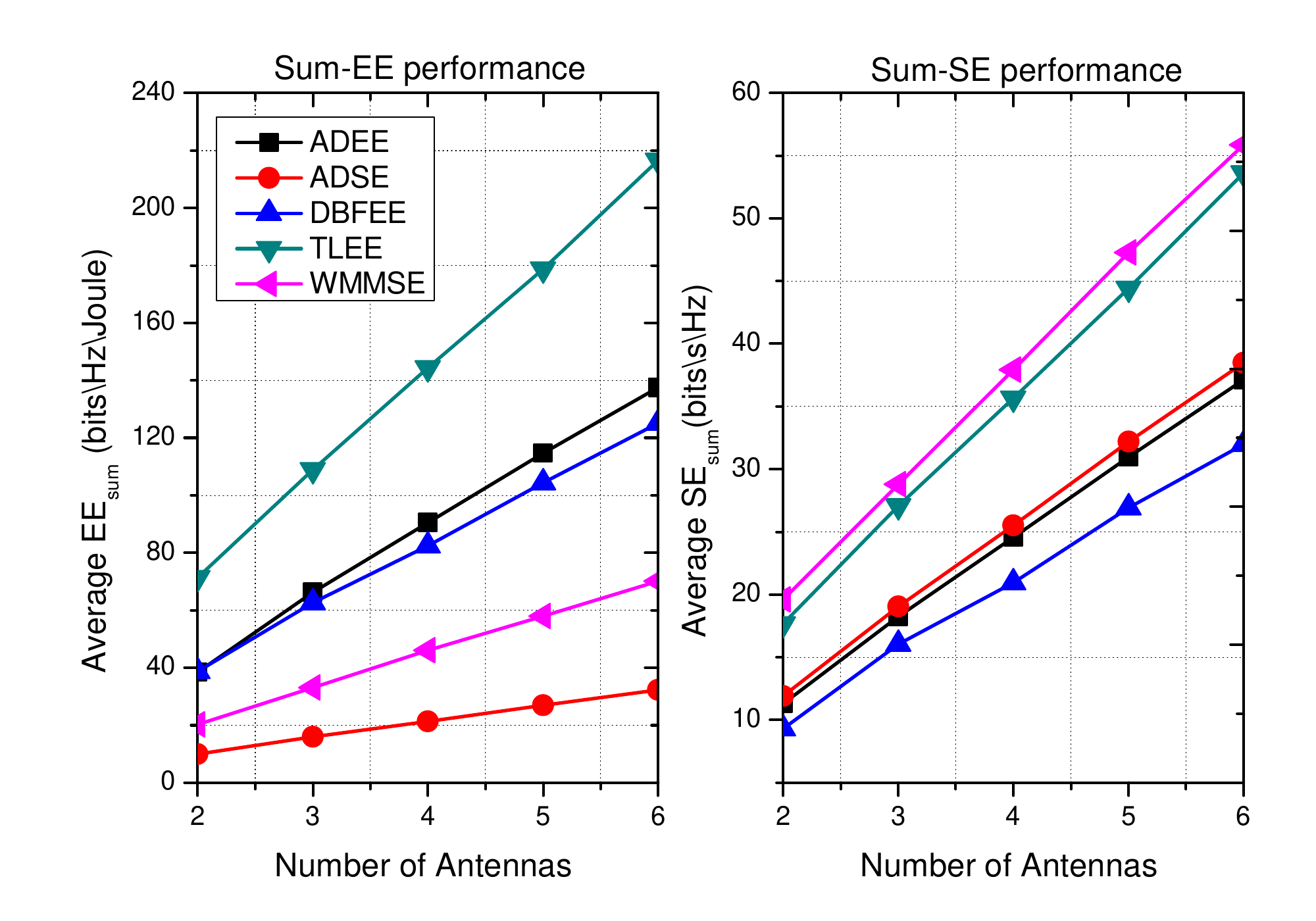}
\caption{Average sum-EE (left) and sum-SE (right) versus the number of antennas for various algorithms.}
\label{Sum-EESE-Antennas}
\end{figure}

Fig.~\ref{Sum-EESE-Antennas} illustrates the average sum-EE (left) and the average sum-SE (right) versus the number of antennas for various algorithms. As expected, increasing the number of antennas at both the transmitters and the receivers leads to a significant increase in both the sum-EE and the sum-SE due to the fact that a larger number of degrees of freedom (DoF) in the spatial domain can be exploited to strengthen the signal power received by the intended receiver while avoiding the interference imposed on the unintended ones. It can be seen from this figure that the ADEE algorithm outperforms the DBF algorithm in terms of the sum-EE and the sum-SE. The performance gain monotonically increases with the number of antennas, meaning the ADEE algorithm uses the spatial resources more effectively. Again, the ADEE algorithm is observed to achieve a significantly higher sum-EE than that achieved by the ADRM algorithm and the performance gain becomes more significant with more antennas. The sum-SE of the ADRM algorithm is slightly higher than that of the ADEE algorithm. As expected, the TLEE algorithm has superior performance over the ADEE algorithm in terms of the sum-EE and sum-SE.

\begin{figure}
\centering
\includegraphics[width=4.5in]{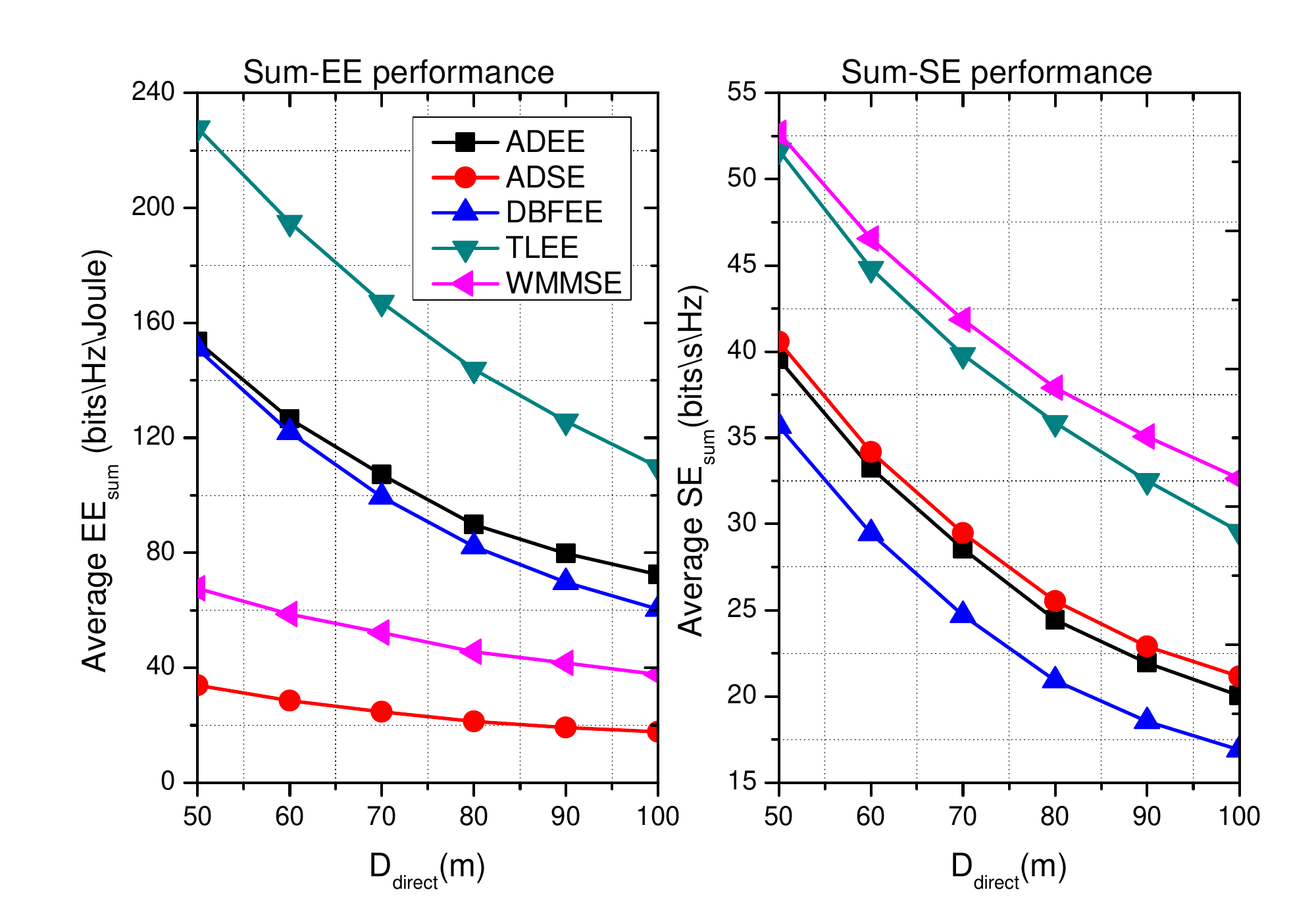}
\caption{Average sum-EE (left) and sum-SE (right) versus the direct-channel distance $D_{\rm{direct}}$ for various algorithms.}
\label{Sum-EESE-distance}
\end{figure}

\subsubsection{Impact of the Direct-Channel Distance on the System Performance}

Fig.~\ref{Sum-EESE-distance} illustrates the impact of increasing the direct-channel distance $D_{\rm{direct}}$ on the sum-EE and sum-SE. It can be seen from this figure that both the sum-EE and sum-SE monotonically decreases upon increasing $D_{\rm{direct}}$. The reason is that increasing $D_{\rm{direct}}$ will degrade the direct-channel gains, and thus reducing the attainable sum-EE and sum-SE. This figure shows much of the same properties as Fig.~\ref{Sum-EESE-Antennas}. For example, the ADEE algorithm outperforms the DBFEE and ADRM algorithms in terms of the sum-EE, and the sum-SE loss is negligible compared with the ADRM algorithm. With much more information exchange overhead and higher computational complexity, the TLEE and WMMSE algorithms have the best sum-EE and sum-SE performance, respectively.

\subsubsection{Impact of the Circuit Power on the System Performance}

\begin{figure}
\centering
\includegraphics[width=4.5in]{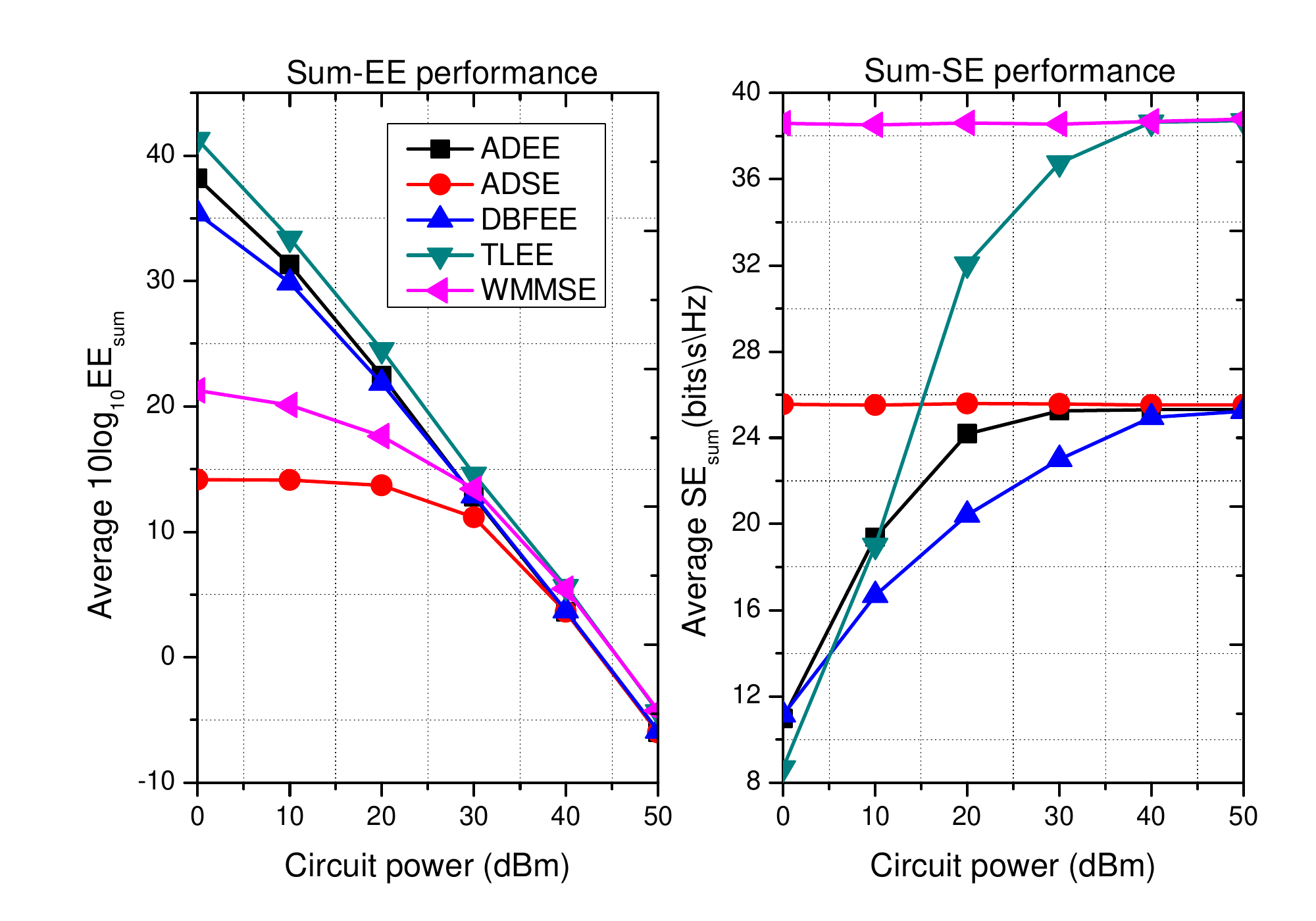}
\caption{Average sum-EE (left) and sum-SE (right) versus the circuit power for various algorithms.}
\label{Sum-EESE-circuit}
\end{figure}

The impact of the circuit power on the attainable sum-EE and sum-SE is characterized in Fig.~\ref{Sum-EESE-circuit}. For illustration purpose, the y-axis is shown by  $10{\rm{log}}_{10}{\rm{EE}}_{\rm{sum}}$. As expected, the sum-EE achieved by all the algorithms decreases with the increase of the circuit power due to the increased power consumption and finally converges to  almost the same value. On the other hand, the corresponding sum-SE of the ADEE algorithm and the DBFEE algorithm  monotonically increases with the increase of circuit power and finally converges to the sum-rate achieved by the ADRM algorithm. This indicates that more circuit power will encourage each link to use more power to achieve a higher SE, which shows the trend in Theorem 4 also holds in the more general case. Note that the similar trend has been observed for the MIMO interference channel in \cite{ChenziJiang-2013}. It is interesting to find that the sum-EE gain provided by the TLEE algorithm over the ADEE algorithm is small and fixed during the overall circuit power consumption regime.

\begin{figure}
\centering
\includegraphics[width=4.5in]{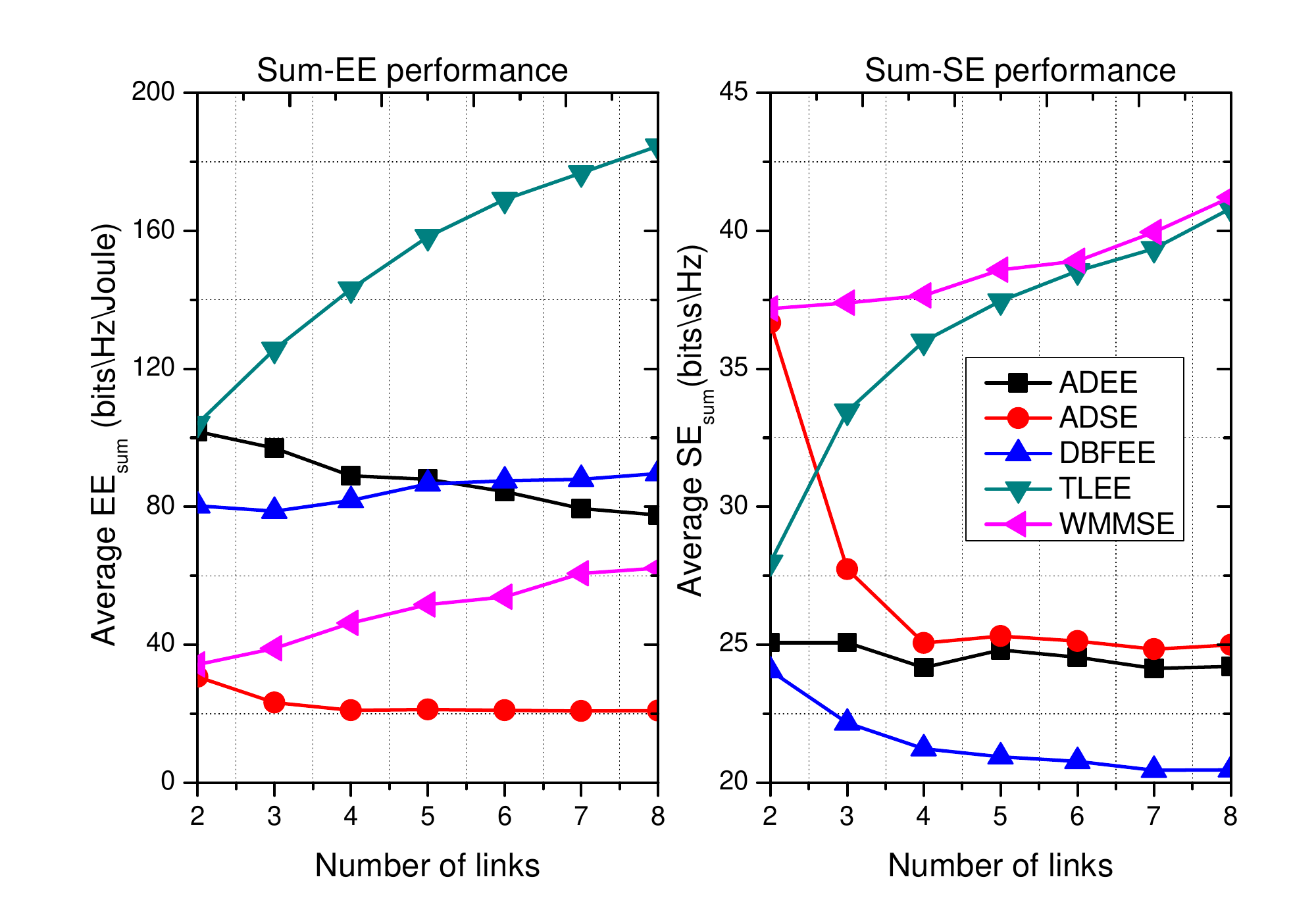}
\caption{Average sum-EE (left) and sum-SE (right) versus the number of links for various algorithms.}
\label{Sum-EESE-K}
\end{figure}

\subsubsection{Impact of number of links on the System Performance}

Finally in Fig.~\ref{Sum-EESE-K}, we investigate the impact of increasing the number of links on the sum-EE and sum-SE performance of different algorithms. It can be seen from this figure that the sum-EE of the ADEE algorithm decreases with the increase of the number of the links. The reason is that when the number of links increases, the interference power received at the receivers increases. To compensate for this negative effects, each transmitter will increase its transmit power. However, this slightly improves the sum-SE but significantly reduces the sum-EE.
To improve the performance of our algorithm for this interference-limited scenario, we may combine our algorithm with some scheduling methods, which is beyond the scope of this paper.
On the other hand, by increasing the number of links, the sum-EE performance of the DBFEE algorithm gradually increases. This can be explained as follows: As the number of links increases, the interference power perceived at different receivers become comparable with each other due to the law of large numbers. As a result, the system will become symmetry, for which the DBFEE algorithm will be suitable.
It can also be observed from this figure that there is a point beyond which the DBFEE algorithm outperforms the ADEE algorithm. However, this superiority comes at the cost of more energy consumption incurred by the information exchange, which is not accounted for in the simulations.
Finally, it is observed that both the sum-EE and sum-SE increases with the number of links due to the multiuser diversity.

\section{Conclusion}\label{conclusion}

In this paper, we have considered a game theoretical formulation of the maximization of the EE on each link, subject to the power constraints, in the MIMO interference channel. We have provided a complete characterization of the game, by showing the existence of the NE and deriving the sufficient conditions for the uniqueness of the NE for the case of large enough maximum transmit power constraint. We have provided a totally asynchronous iterative distributed algorithm, named the ADEE algorithm, to reach the NE of this game. The ADEE algorithm has three advantages: no rigid scheduling in the updates of the players is required, the synchronization requirement of both sequential and simultaneous algorithms can be removed and information exchange among the players is not necessary, which make the proposed algorithm appealing to implement in practice. Furthermore, the sufficient conditions guaranteeing the convergence of the algorithm have been provided. Interestingly, our sufficient condition does not depend on the scheduling of the links. Extensive simulation results have shown that the ADEE algorithm performs better than the DBFEE and ADRM algorithms in terms of the sum-EE performance, and is comparable to the ADRM algorithm in terms of the sum-SE performance.

\numberwithin{equation}{section}
\begin{appendices}

\section{Proof of Lemma 1}\label{prooflemma1}
Before proving Lemma 1, the following definition is given.

\itshape \textbf{Definition 2  \cite{Boyd2004}:}  \upshape A real-valued differentiable function  $f({\bf{X}}):{\bf{X}} \in \mathbb{S}_ + ^{M \times M}$, is (strictly) quasiconcave on $\mathbb{S}_ + ^{M \times M}$  if its sublevel sets ${S_\alpha } = \{ {\bf{X}} \in \mathbb{S}_ + ^{M \times M}|f({\bf{X}}) \ge \alpha \}$, for $\alpha  \in {\bf{R}}$, are (strictly) convex. \hfill $\Box$

For simplicity, we omit the dependency of  ${\rm{EE}}_k({{\bf{Q}}_k},{{\bf{Q}}_{ - k}})$ and $C_k({{\bf{Q}}_k},{{\bf{Q}}_{ - k}})$ on   ${{\bf{Q}}_{ - k}}$. Define two functions ${f_1}({{\bf{Q}}_k}) = {C_k}({{\bf{Q}}_k})$ and ${f_2}({{\bf{Q}}_k}) = {\rm{tr(}}{{\bf{Q}}_k}{\rm{) + }}{P_{\rm{C}}}$, and define $f({{\bf{Q}}_k})\buildrel \Delta \over ={\rm{EE}}_k({{\bf{Q}}_k}) = \frac{{{f_1}({{\bf{Q}}_k})}}{{{f_2}({{\bf{Q}}_k})}}$.

Now, we give the proof of the lemma.  Since  ${f_2}({{\bf{Q}}_k}) > 0$, ${S_\alpha }$  is equivalent to
\begin{equation}\label{equalto}
    {S_\alpha } = \{ {\bf{Q}}_k \in \mathbb{S}_ + ^{M \times M}|{f_1}({{\bf{Q}}_k}) - \alpha {f_2}({{\bf{Q}}_k}) \ge 0\}.
\end{equation}
Since  ${f_2}({{\bf{Q}}_k})$  is affine in  ${{\bf{Q}}_k}$, ${S_\alpha }$ is (strictly) convex for any given  $\alpha $ if  ${f_1}({{\bf{Q}}_k})$  is (strictly) concave in  ${{\bf{Q}}_k}$. In the following, we will prove ${f_1}({{\bf{Q}}_k})$  is concave w.r.t. ${{\bf{Q}}_k}$ for any channel matrix ${{\bf{H}}_{k,k}}$ and is strictly concave w.r.t. ${{\bf{Q}}_k}$ if ${{\bf{H}}_{k,k}}$ is full column rank.

To this end, we adopt the technique in \cite{Sigen-2003}.  We consider the convex combination of two different feasible  ${{{\bf{ X}}}_k} \in {{{\cal W}}_k}$ and ${{{\bf{ Z}}}_k} \in {{{\cal W}}_k}$, which is
\begin{equation}\label{ddss}
\begin{array}{c}
{{{\bf{ Q}}}_k} = t{{{\bf{ Z}}}_k} + (1 - t){{{\bf{ X}}}_k} = {{{\bf{ X}}}_k} + t{{{\bf{ Y}}}_k},
\end{array}
\end{equation}
where $0 \leq t \leq 1$, and ${{{\bf{ Y}}}_k} = {{{\bf{ Z}}}_k} - {{{\bf{X}}}_k}$, which is a non-zero Hermitian matrix. Obviously, ${{{\bf{ Q}}}_k} \in {{ {\cal W}}_k}$. Then  $f_1({{{\bf{ Q}}}_k})$ is (strictly) concave w.r.t. ${{{\bf{ Q}}}_k}$ if and only if $({{{{\rm{d}}^2}} \mathord{\left/
 {\vphantom {{{{\rm{d}}^2}} {{\rm{d}}{t^2})}}} \right.
 \kern-\nulldelimiterspace} {{\rm{d}}{t^2})}}f_1({{{\bf{ Q}}}_k}) \leq 0$ $(({{{{\rm{d}}^2}} \mathord{\left/
 {\vphantom {{{{\rm{d}}^2}} {{\rm{d}}{t^2})}}} \right.
 \kern-\nulldelimiterspace} {{\rm{d}}{t^2})}}f_1({{{\bf{ Q}}}_k}) < 0)$ for any feasible ${{{\bf{ X}}}_k}$ and ${{{\bf{ Z}}}_k}$. Denote ${{\bf{A}}_k} = {\bf{ H}}_{k,k}^H{\bf{ R}}_k^{ - 1}{{{\bf{ H}}}_{k,k}}$, the second derivative of $f_1({{{\bf{ Q}}}_k})$ w.r.t. $t$ can be calculated as
 \begin{equation}\label{twoderivative}
   \frac{{{{\rm{d}}^2}f_1({{{\bf{ Q}}}_k})}}{{{\rm{d}}{t^2}}} =  - \frac{1}{{\ln 2}}{\rm{tr}}\left( {{{\bf{A}}_k}{{{\bf{ Y}}}_k}{{\left( {{\bf{I}} + {{\bf{A}}_k}{{{\bf{ Q}}}_k}} \right)}^{ - 1}}{{\bf{A}}_k}{{{\bf{ Y}}}_k}{{\left( {{\bf{I}} + {{\bf{A}}_k}{{{\bf{ Q}}}_k}} \right)}^{ - 1}}} \right).
 \end{equation}
 Since ${\left( {{\bf{I}} + {{\bf{A}}_k}{{{\bf{ Q}}}_k}} \right)^{ - 1}} \in \mathbb{S}_{ +  + }^{M \times M}$, it is possible to  write ${\left( {{\bf{I}} + {{\bf{A}}_k}{{{\bf{ Q}}}_k}} \right)^{ - 1}} = {{\bf{D}}_k}{\bf{D}}_k^H$ with ${\bf{D}}_k^H \in \mathbb{S}_{ +  + }^{M \times M}$.  Then, it is seen that
 \begin{equation}\label{finalproof}
\frac{{{{\rm{d}}^2}f_1({{{\bf{ Q}}}_k})}}{{{\rm{d}}{t^2}}} =  - \frac{1}{{\ln 2}}{\rm{tr}}\left( {\left( {{\bf{D}}_k^H{{\bf{A}}_k}{{{\bf{ Y}}}_k}{{\bf{D}}_k}} \right)\left( {{\bf{D}}_k^H{{\bf{A}}_k}{{{\bf{ Y}}}_k}{{\bf{D}}_k}} \right)} \right) \leq 0.
 \end{equation}
Furthermore, when ${{{\bf{ H}}}_{k,k}}$ is a full column-rank matrix,  ${{{\bf{A}}_k}{{{\bf{Y}}}_k}}$ is always non-zero since ${{{{\bf{ Y}}}_k}}$ is a non-zero matrix. Then $({{{{\rm{d}}^2}} \mathord{\left/
 {\vphantom {{{{\rm{d}}^2}} {{\rm{d}}{t^2})}}} \right.
 \kern-\nulldelimiterspace} {{\rm{d}}{t^2})}}{f_1}({{{\bf{Q}}}_k}) < 0$ always holds. Hence, the proof is completed.

\section{Proof of Theorem 1}\label{proofTheorem1}
We prove the existence of NE for game ${\cal G}$ using the following well-known game theory result.

\itshape \textbf{Lemma 5 \cite{Osborne-1994} :}  \upshape The strategic noncooperative game ${\cal G} = \left\{ {{\Psi _K},{{{\rm{\{ }}{{\cal W}_k}{\rm{\} }}}_{k \in {\Psi _K}}},{{\{ {U_k}\} }_{k \in {\Psi _K}}}} \right\}$  admits at least one NE if, for all  $k \in {\Psi _K}$:1) the set ${{\cal W}_k}$  is a nonempty compact convex subset of a Euclidean space, and 2) the payoff function ${U_k}$  is  quasiconcave on  ${{\cal W}_k}$. \hfill $\Box$

It can be easily verified that game ${\cal G}$  satisfies the two conditions required by Lemma 5 according to Lemma 1 and the convexity of the admissible power set.

\section{Proof of Theorem 2}\label{proofTheorem2}

Before proving Theorem 2, we provide three lemmas \cite{Horn-2012} that will be used in our derivations.

\itshape \textbf{Lemma 6:}  \upshape
Given matrices ${\bf{A}} \in \mathbb{C}^{m \times n}, {\bf{B}} \in \mathbb{C}^{n \times l}$,  the relation
${\left\| {{\bf{AB}}} \right\|_2} \le {\left\| {\bf{A}} \right\|_2}{\left\| {\bf{B}} \right\|_2}$ holds.\hfill $\Box$

\itshape \textbf{Lemma 7:}  \upshape If ${\bf{A}},{\bf{B}} \in \mathbb{S}_{ +  + }^{n \times n}$, then we have
$\rho ({\bf{AB}}) \le \rho ({\bf{A}})\rho ({\bf{B}})$.\hfill $\Box$

\itshape \textbf{Lemma 8:}  \upshape
Given matrix ${\bf{A}} \in \mathbb{C}^{n \times n}$ with eigenvalues $\left\{ {{\lambda _i},i = 1, \cdots ,n} \right\}$ and eigenvectors  $\left\{ {{{\bf{x}}_i},i = 1, \cdots ,n} \right\}$, and matrix ${\bf{B}} \in \mathbb{C}^{m \times m}$ with eigenvalues $\left\{ {{\mu _i},i = 1, \cdots ,n} \right\}$ and  eigenvectors $\left\{ {{{\bf{y}}_i},i = 1, \cdots ,n} \right\}$, the eigenvalues of ${\bf{A}} \otimes {\bf{B}}$ are given by $\left\{ {{\lambda _i}{\mu _j},i = 1, \cdots ,n;j = 1, \cdots ,m} \right\}$ and the corresponding eigenvectors are given by ${{{\bf{x}}_i} \otimes {{\bf{y}}_j},i = 1, \cdots ,n;j = 1, \cdots ,m}$.\hfill $\Box$

Based on the above results, we now proceed to prove Theorem 2. Our main idea lies in showing that under condition (\ref{uniquecondition}),  condition (\ref{lemma2}) in Lemma 3 will hold. To this end, we will employ the mean-value theorem for complex matrix-valued functions stated in \cite{Scutari-2009}. That is, for any two different points ${\bf{Q}}_{ - k}^{(1)}$  and  ${\bf{Q}}_{ - k}^{(2)}$, there exists some $t \in (0,1)$  such that
\begin{equation}\label{mean-value}
    {\left\| {{{\bf{F}}_k}({\bf{Q}}_{ - k}^{(1)})\! - \!{{\bf{F}}_k}({\bf{Q}}_{ - k}^{(2)})} \right\|_F}\! \le \!{\left\| {{{\bf{D}}_{{{\bf{Q}}_{ - k}}}}{{\bf{F}}_k}({\bf{\Delta}} )} \right\|_2}{\left\| {{\bf{Q}}_{ - k}^{(1)}\! -\! {\bf{Q}}_{ - k}^{(2)}} \right\|_F},
\end{equation}
where  ${\bf{\Delta}}= t{\bf{Q}}_{ - k}^{(1)} + (1 - t){\bf{Q}}_{ - k}^{(2)}$  and ${{\bf{D}}_{{{\bf{Q}}_{ - k}}}}{{\bf{F}}_k}({\bf{\Delta}} )$ represents the Jacobian matrix of function ${{\bf{F}}_k}({{\bf{Q}}_{ - k}})$ w.r.t. ${{\bf{Q}}_{ - k}}$, evaluated at ${{\bf{Q}}_{ - k}} = {\bf{\Delta}} $ \cite{Hjorungnes-2007}.

Let ${{\bf{D}}_{{{\bf{R}}_k}}}{{\bf{F}}_k}({\bf{\Delta}})$ be the Jacobian matrix of function ${{\bf{F}}_k}({{\bf{Q}}_{ - k}})$ w.r.t. ${{\bf{R}}_k}$,  evaluated at ${{\bf{R}}_k} = {{\bf{R}}_k}({\bf{\Delta}})$. Let ${{\bf{D}}_{{{\bf{Q}}_{ - k}}}}{{\bf{R}}_k}({\bf{\Delta}})$ be the Jacobian matrix of function ${{\bf{R}}_k}({{\bf{Q}}_{ - k}})$ w.r.t. ${{\bf{Q}}_{ - k}}$, evaluated at ${{\bf{Q}}_{ - k}} = {\bf{\Delta}}$. Then, we have:
\begin{eqnarray}
\!\!\!\!{\left\| {{{\bf{F}}_k}({\bf{Q}}_{ - k}^{(1)}) - {{\bf{F}}_k}({\bf{Q}}_{ - k}^{(2)})} \right\|_F}
&\le& \! \!\! \!{\left\| {{{\bf{D}}_{{{\bf{R}}_k}}}{{\bf{F}}_k}({\bf{\Delta}})\cdot{{\bf{D}}_{{{\bf{Q}}_{ - k}}}}{{\bf{R}}_k}}({\bf{\Delta}}) \right\|_2}{\left\| {{\bf{Q}}_{ - k}^{(1)} - {\bf{Q}}_{ - k}^{(2)}} \right\|_F}\label{chainrule}\\
 &\le& \!\!\! \!{\left\| {{{\bf{D}}_{{{\bf{R}}_k}}}{{\bf{F}}_k}}({\bf{\Delta}}) \right\|_2}{\left\| {{{\bf{D}}_{{{\bf{Q}}_{ - k}}}}{{\bf{R}}_k}({\bf{\Delta}})} \right\|_2}{\left\| {{\bf{Q}}_{ - k}^{(1)} - {\bf{Q}}_{ - k}^{(2)}} \right\|_F},\label{compatibility}
\end{eqnarray}
where (\ref{chainrule}) follows from the chain rule and (\ref{compatibility}) follows from Lemma 6.

Our goal is to obtain the upper bound of ${\left\| {{{\bf{D}}_{{{\bf{R}}_k}}}{{\bf{F}}_k}({\bf{\Delta}})} \right\|_2}{\left\| {{{\bf{D}}_{{{\bf{Q}}_{ - k}}}}{{\bf{R}}_k}({\bf{\Delta}})} \right\|_2}$. If this upper bound is less than $\sqrt{{1 \mathord{\left/
 {\vphantom {1 {(K - 1)}}} \right.
 \kern-\nulldelimiterspace} {(K - 1)}}}$, then Condition (\ref{lemma2}) is satisfied. In the following, we derive the upper bounds of ${\left\| {{{\bf{D}}_{{{\bf{R}}_k}}}{{\bf{F}}_k}({\bf{\Delta}})} \right\|_2}$ and ${\left\| {{{\bf{D}}_{{{\bf{Q}}_{ - k}}}}{{\bf{R}}_k}({\bf{\Delta}})} \right\|_2}$, respectively.

\subsection {\textbf{The upper bound of} ${\left\| {{{\bf{D}}_{{{\bf{Q}}_{ - k}}}}{{\bf{R}}_k}({\bf{\Delta}})} \right\|_2}$} To derive the upper bound of ${\left\| {{{\bf{D}}_{{{\bf{Q}}_{ - k}}}}{{\bf{R}}_k}({\bf{\Delta}})} \right\|_2}$, we should  obtain the expression of ${{{\bf{D}}_{{{\bf{Q}}_{ - k}}}}{{\bf{R}}_k}({\bf{\Delta}})}$. To this end, the expression of the Jacobian matrix ${{{\bf{D}}_{{{\bf{Q}}_{ - k}}}}{{\bf{R}}_k}}$, which is a function of ${{\bf{Q}}_{ - k}}$, should be obtained firstly. Then ${{{\bf{D}}_{{{\bf{Q}}_{ - k}}}}{{\bf{R}}_k}({\bf{\Delta}})}$ can be obtained by inserting ${{\bf{Q}}_{ - k}} = {\bf{\Delta}}$ into ${{{\bf{D}}_{{{\bf{Q}}_{ - k}}}}{{\bf{R}}_k}}$.

The Jacobian matrix ${{{\bf{D}}_{{{\bf{Q}}_{ - k}}}}{{\bf{R}}_k}}$ can be computed using the three-step procedure in \cite{Hjorungnes-2007}. Specifically, we compute firstly the differential of  ${\bf{R}}_k$ and then the Jacobian matrix. Function ${\bf{R}}_k$  is differentiable at  ${{\bf{Q}}_{ - k}}$, with differential given by
${\rm{d}}{\bf{R}}_k = \sum\limits_{j \ne k} {{{\bf{H}}_{j,k}}{\rm{d}}{{\bf{Q}}_j}{\bf{H}}_{j,k}^H}$.
By vectorizing ${\rm{d}}{\bf{R}}_k$  and using the equality ${\rm{vec}}({\bf{ABC}}) = ({{\bf{C}}^T} \otimes {\bf{A}}){\rm{vec(}}{\bf{B}}{\rm{)}}$ \cite{Horn-2012}, we obtain
\begin{equation}\label{vec}
{\rm{d}}{\rm{vec}}{{\bf{R}}_k}\!\! =\!\! \left[\! {{\bf{H}}_{1,k}^*\! \otimes\! {{\bf{H}}_{1,k}},\! \cdots ,{\bf{H}}_{k - 1,k}^* \! \otimes\! {{\bf{H}}_{k - 1,k}},\!{\bf{H}}_{k + 1,k}^* \!\otimes\! {{\bf{H}}_{k + 1,k}},\! \cdots ,\!{\bf{H}}_{K,k}^* \otimes {{\bf{H}}_{K,k}}} \right]{\rm{d}}{\rm{vec}}({{\bf{Q}}_{ - k}}).
\end{equation}
Using the identification rule in \cite{Hjorungnes-2007}, we can obtain
${{\bf{D}}_{{{\bf{Q}}_{ - k}}}}{\bf{R}}_k$ given in (\ref{Jociab}) in Theorem 2. Note that ${{\bf{D}}_{{{\bf{Q}}_{ - k}}}}{\bf{R}}_k$ does not depend on ${{\bf{Q}}_{ - k}}$. Hence, we have
${{\bf{D}}_{{{\bf{Q}}_{ - k}}}}{\bf{R}}_k({\bf{\Delta}})={{\bf{D}}_{{{\bf{Q}}_{ - k}}}}{\bf{R}}_k$.

\subsection {\textbf{The upper bound of} ${\left\| {{{\bf{D}}_{{{\bf{R}}_k}}}{{\bf{F}}_k}}({\bf{\Delta}}) \right\|_2}$}
Similarly, to obtain the upper bound of  ${\left\| {{{\bf{D}}_{{{\bf{R}}_k}}}{{\bf{F}}_k}}({\bf{\Delta}}) \right\|_2}$, we should  obtain the expression of ${{{\bf{D}}_{{{\bf{R}}_k}}}{{\bf{F}}_k}}$, which is a function of ${{\bf{Q}}_{ - k}}$. Then the expression of ${{{\bf{D}}_{{{\bf{R}}_k}}}{{\bf{F}}_k}}({\bf{\Delta}})$ can be obtained by inserting ${{\bf{Q}}_{ - k}} = {\bf{\Delta}}$ into it.

According to Lemma 1, the EE function ${\rm{EE}}_k({{\bf{Q}}_k})$ is strictly quasiconcave on $\mathbb{S}_ + ^{M \times M}$ since ${{{\bf{H}}_{k,k}}}$ is assumed to be full column rank. Also, the maximum transmit power is assumed to be very large. Then, for given ${{\bf{Q}}_{ - k}}$ (and thus ${{\bf{R}}_k}$), the gradient of ${\rm{EE}}_k({{\bf{Q}}_k})$ w.r.t. ${{\bf{Q}}_k}$, evaluated at  ${{\bf{Q}}_k}{\rm{ = }}{{\bf{F}}_k}$, must be zero \cite{Wolfstetter-1999}.
 Thus, using ${{\partial \left| {{\bf{BXC}}} \right|} \mathord{\left/
 {\vphantom {{\partial \left| {{\bf{BXC}}} \right|} {\partial {\bf{X}} = }}} \right.
 \kern-\nulldelimiterspace} {\partial {\bf{X}} = }}\left| {{\bf{BXC}}} \right|{\left[ {{\bf{C}}{{({\bf{BXC}})}^{ - 1}}{\bf{B}}} \right]^T}$ \cite{Magnus-1988}, we have
\begin{equation}\label{Gfunction}
 {\bf{G}}({{\bf{R}}_k},{{\bf{F}}_k})\! \buildrel \Delta \over = \!{\bf{H}}_{k,k}^H{({{\bf{R}}_k} + {{\bf{H}}_{k,k}}{{\bf{F}}_k}{\bf{H}}_{k,k}^H)^{ - 1}}{{\bf{H}}_{k,k}}\! -\! {\alpha _k}{\bf{I}}\! = \!{\bf{0}}.
\end{equation}
where ${\alpha _k} = {{{C_k}({{\bf{Q}}_k},{{\bf{Q}}_{ - k}})} \mathord{\left/
 {\vphantom {{{C_k}} {({P_k} + {P_C})}}} \right.
 \kern-\nulldelimiterspace} {({P_k} + {P_C})}}$.
Since function ${\rm{EE}}_k({{\bf{Q}}_k})$ is strictly quasiconcave in ${{\bf{Q}}_k}$  for given  ${{\bf{R}}_k}$, there exists a unique globally optimal solution ${{\bf{F}}_k}$  that satisfies (\ref{Gfunction}) \cite{Wolfstetter-1999}. Hence, equation (\ref{Gfunction}) defines an implicit function \cite{Jittorntrum-1978}. Taking the derivative of  (\ref{Gfunction}) w.r.t. ${{\bf{R}}_k}$ and using the chain rule, we have \footnote{For simplicity, the dependency of function ${\bf{G}}({{\bf{R}}_k},{{\bf{F}}_k})$ on both ${{\bf{R}}_k}$ and ${{\bf{F}}_k}$ is omitted.}
\begin{equation}\label{Differentialbothsides}
    {{\bf{D}}_{{{\bf{R}}_k}}}{\bf{G}} + {{\bf{D}}_{{{\bf{F}}_k}}}{\bf{G}}\cdot{{\bf{D}}_{{{\bf{R}}_k}}}{{\bf{F}}_k} = {\bf{0}}.
\end{equation}

Now we first obtain the Jacobian matrices  ${{\bf{D}}_{{{\bf{R}}_k}}}{\bf{G}}$ and ${{\bf{D}}_{{{\bf{F}}_k}}}{\bf{G}}$, then ${{{\bf{D}}_{{{\bf{R}}_k}}}{{\bf{F}}_k}}$ can be solved from (\ref{Differentialbothsides}). We again use the three-step procedure in \cite{Hjorungnes-2007} to compute ${{\bf{D}}_{{{\bf{R}}_k}}}{\bf{G}}$ and ${{\bf{D}}_{{{\bf{F}}_k}}}{\bf{G}}$. Function ${\bf{G}}$ is differentiable w.r.t. both ${{\bf{R}}_k}$ and ${{\bf{F}}_k}$, with the differential given by
\begin{equation}\label{differentialwrtrf}
    {\rm{d}}{\bf{G}} =  - {\bf{H}}_{k,k}^H{{\bf{J}}_k}{\rm{d}}{{\bf{R}}_k}{{\bf{J}}_k}{{\bf{H}}_{k,k}} - {\bf{H}}_{k,k}^H{{\bf{J}}_k}{{\bf{H}}_{k,k}}{\rm{d}}{{\bf{F}}_k}{\bf{H}}_{k,k}^H{{\bf{J}}_k}{{\bf{H}}_{k,k}}.
\end{equation}
where ${{\bf{J}}_k} = {\left( {{{\bf{R}}_k} + {{\bf{H}}_{k,k}}{{\bf{F}}_k}{\bf{H}}_{k,k}^H} \right)^{ - 1}}$ and we used ${\rm{d}}{{\bf{X}}^{ - 1}} =  - {{\bf{X}}^{ - 1}}{\rm{d}}{\bf{X}}{{\bf{X}}^{ - 1}}$ \cite{Hjorungnes-2007}. Then by vectorizing both sides of (\ref{differentialwrtrf}) and using the identification rule in \cite{Hjorungnes-2007}, we can obtain the Jacobian matrices  ${{\bf{D}}_{{{\bf{R}}_k}}}{\bf{G}} =  - {{\bf{U}}_k}$ and ${{\bf{D}}_{{{\bf{F}}_k}}}{\bf{G}} =  - {{\bf{V}}_k}$ , with ${{\bf{U}}_k}  = {\bf{H}}_{k,k}^T{\bf{J}}_k^T \otimes {\bf{H}}_{k,k}^H{{\bf{J}}_k}$ and  ${{\bf{V}}_k}  = {\bf{H}}_{k,k}^T{\bf{J}}_k^T{\bf{H}}_{k,k}^* \otimes {\bf{H}}_{k,k}^H{{\bf{J}}_k}{{\bf{H}}_{k,k}}$. Note that ${{\bf{V}}_k}$ is a Hermitian matrix. Since the channel matrix ${{\bf{H}}_{k,k}}$ is assumed to be full column rank and ${{\bf{J}}_k}$ is nonsingular, ${\bf{H}}_{k,k}^H{{\bf{J}}_k}{{\bf{H}}_{k,k}}$  is nonsingular. Then ${{\bf{V}}_k}$  is also nonsingular. This is because \cite{Magnus-1988}
\begin{equation}\label{vecinverse}
  {\left( {{\bf{A}} \otimes {\bf{B}}} \right)^{ - 1}} = {{\bf{A}}^{ - 1}} \otimes {{\bf{B}}^{ - 1}}, {\rm{for\  all \ nonsingular}}\  {\bf{A,B}}.
\end{equation}
Hence, ${{\bf{D}}_{{{\bf{R}}_k}}}{{\bf{F}}_k}$  can be solved from (\ref{Differentialbothsides}):
\begin{equation}\label{JocabiFR}
    {{\bf{D}}_{{{\bf{R}}_k}}}{{\bf{F}}_k} =  - {\bf{V}}_k^{ - 1}{{\bf{U}}_k}.
\end{equation}
Thus, ${{\bf{D}}_{{{\bf{R}}_k}}}{{\bf{F}}_k}({\bf{\Delta}})$ can be obtained by inserting ${{\bf{Q}}_{ - k}} = {\bf{\Delta}}$ into (\ref{JocabiFR}):
\begin{equation}\label{bb}
  {{\bf{D}}_{{{\bf{R}}_k}}}{{\bf{F}}_k}({\bf{\Delta}}) =  - {\bf{V}}_k^{ - 1}({\bf{\Delta}}){{\bf{U}}_k}({\bf{\Delta}}).
\end{equation}
Then, we have
\begin{equation}\label{SpectalFR}
    \left\| {{{\bf{D}}_{{{\bf{R}}_k}}}{{\bf{F}}_k}({\bf{\Delta}})} \right\|_2^2 = \rho ({\bf{U}}_k^H({\bf{\Delta}}){\bf{V}}_k^{ - 2}({\bf{\Delta}}){{\bf{U}}_k}({\bf{\Delta}})) \le \rho ({\bf{V}}_k^{ - 2}({\bf{\Delta}}))\rho ({{\bf{U}}_k}({\bf{\Delta}}){\bf{U}}_k^H({\bf{\Delta}})),
\end{equation}
where the inequality follows from Lemma 7. We now aim to obtain the upper bound of $\rho ({\bf{V}}_k^{ - 2}({\bf{\Delta}}))$  and   $\rho ({{\bf{U}}_k}({\bf{\Delta}}){\bf{U}}_k^H({\bf{\Delta}}))$, respectively.

\subsubsection {The upper bound of $\rho ({\bf{V}}_k^{ - 2}({\bf{\Delta}}))$} The upper bound of $\rho ({\bf{V}}_k^{ - 2}({\bf{\Delta}}))$  can be obtained as follows:
\begin{eqnarray}
&&\!\!\!\rho ({\bf{V}}_k^{ - 2}({\bf{\Delta}}))\\
 &=&\!\! \rho \left( {{{\left( {{\bf{H}}_{k,k}^T{\bf{J}}_k^T({\bf{\Delta}}){\bf{H}}_{k,k}^* \otimes {\bf{H}}_{k,k}^H{{\bf{J}}_k}({\bf{\Delta}}){{\bf{H}}_{k,k}}} \right)}^{ - 2}}} \right)\label{54}\\
 &=& \!\!\rho \left( {{{\left( {{\bf{H}}_{k,k}^T{\bf{J}}_k^T({\bf{\Delta}}){\bf{H}}_{k,k}^*} \right)}^{ - 2}} \otimes {{\left( {{\bf{H}}_{k,k}^H{{\bf{J}}_k}({\bf{\Delta}}){{\bf{H}}_{k,k}}} \right)}^{ - 2}}} \right)\label{55}\\
 &=&\!\!{\left( {\rho \left( {{{\left( {{\bf{H}}_{k,k}^H{{\bf{J}}_k}({\bf{\Delta}}){{\bf{H}}_{k,k}}} \right)}^{ - 2}}} \right)} \right)^2}\label{56}\\
 &=&\!\!{\left( {{\lambda _{\min }}\left( {{\bf{H}}_{k,k}^H{{\bf{J}}_k}({\bf{\Delta}}){{\bf{H}}_{k,k}}} \right)} \right)^{ - 4}}\label{57}\\
 &\le&\! {\left(\!\! {{\lambda _{\min }}\left(\!\! {{\bf{H}}_{k,k}^H{{\left( {\!\!{\bf{I}} + {P_T}\sum\limits_{i = 1}^K {{{\bf{H}}_{i,k}}{\bf{H}}_{i,k}^H} }\!\! \right)}^{ - 1}}\!\!\!{{\bf{H}}_{k,k}}}\!\! \right)}\!\! \right)^{ - 4}}\label{58}
\end{eqnarray}
where ${{\bf{J}}_k}({\bf{\Delta}})$ is obtained by inserting ${{\bf{Q}}_{ - k}} = {\bf{\Delta}}$ into ${{\bf{J}}_k}$, (\ref{55}) follows from (\ref{vecinverse}) and the equality $\left( {{\bf{A}} \otimes {\bf{B}}} \right)\left( {{\bf{C}} \otimes {\bf{D}}} \right) = {\bf{AC}} \otimes {\bf{BD}}$ \cite{Magnus-1988}, (\ref{56}) follows from Lemma 8,  (\ref{57}) results from the fact that  ${\bf{H}}_k^H{{\bf{J}}_k}({\bf{\Delta}}){{\bf{H}}_k} \in \mathbb{S}_{ +  + }^{M \times M}$, the last inequality results from
${\lambda _{\min }}({{\bf{A}}^H}{\bf{BA}}) \ge {\lambda _{\min }}({{\bf{A}}^H}{\bf{CA}}){\rm{ \  for\  all\   }}{\bf{B}}\succeq{\bf{C}}\succeq{\bf{0}}$ and the following relations
\begin{align}
{{\bf{J}}_k}& = {\left( {{\bf{I}} + \sum\limits_{i \ne k} {{{\bf{H}}_{i,k}}{\bf{\Delta}}{\bf{H}}_{i,k}^H}  + {{\bf{H}}_{k,k}}{{\bf{F}}_k}(\Delta ){\bf{H}}_{k,k}^H} \right)^{ - 1}}\label{13}\\
&\succeq{\left( {{\bf{I}} + {\rm{tr}}\left( {{{\bf{F}}_k}(\Delta )} \right){{\bf{H}}_{k,k}}{\bf{H}}_{k,k}^H + \sum\limits_{i \ne k} {{\rm{tr}}\left( {\bf{\Delta}} \right){{\bf{H}}_{i,k}}{\bf{H}}_{i,k}^H} } \right)^{ - 1}}\label{14}\\
&\succeq{\left( {{\bf{I}} + {P_T}\sum\limits_{i = 1}^K {{{\bf{H}}_{i,k}}{\bf{H}}_{i,k}^H} } \right)^{ - 1}}\label{15}.
\end{align}

\subsubsection {The upper bound of $\rho ({{\bf{U}}_k({\bf{\Delta}})}{\bf{U}}_k^H({\bf{\Delta}}))$}
The upper bound of  $\rho ({{\bf{U}}_k({\bf{\Delta}})}{\bf{U}}_k^H({\bf{\Delta}}))$ can be obtained as follows:
\begin{eqnarray}
\!\!\rho ({{\bf{U}}_k({\bf{\Delta}})}{\bf{U}}_k^H({\bf{\Delta}}))\!\!\!\!&=&\!\!\!\! \rho \left( {\left( {{\bf{H}}_{k,k}^T{\bf{J}}_k^T({\bf{\Delta}})) \otimes {\bf{H}}_{k,k}^H{{\bf{J}}_k}({\bf{\Delta}}))} \right)\!\!\left( {{\bf{J}}_k^*({\bf{\Delta}})){\bf{H}}_{k,k}^* \otimes {\bf{J}}_k^H({\bf{\Delta}})){{\bf{H}}_{k,k}}} \right)} \right)\\
 &=&\!\! \!\!\rho \left( {{\bf{H}}_{k,k}^T{\bf{J}}_k^T({\bf{\Delta}}){\bf{J}}_k^*({\bf{\Delta}}){\bf{H}}_{k,k}^* \otimes {\bf{H}}_{k,k}^H{{\bf{J}}_k}({\bf{\Delta}}){\bf{J}}_k^H({\bf{\Delta}}){{\bf{H}}_{k,k}}} \right)\label{62}\\
 &=&\!\! \!\!{\left( {\rho \left( {{\bf{H}}_{k,k}^H{\bf{J}}_k^2({\bf{\Delta}}){{\bf{H}}_{k,k}}} \right)} \right)^2}\label{63}\\
 &\le&\!\!\!\! {\left( {\rho \left( {{\bf{H}}_{k,k}^H{{\bf{H}}_{k,k}}} \right)} \right)^2}\label{61}
\end{eqnarray}
where in (\ref{62}) we use  $\left( {{\bf{A}} \otimes {\bf{B}}} \right)\left( {{\bf{C}} \otimes {\bf{D}}} \right) = {\bf{AC}} \otimes {\bf{BD}}$ \cite{Magnus-1988} , (\ref{63}) follows from Lemma 8, and (\ref{61})  follows from ${{\bf{J}}_k({\bf{\Delta}})}\preceq{\bf{I}}$.

Combining (\ref{SpectalFR}) with  (\ref{58}) and (\ref{61}), we obtain

\begin{equation}\label{second}
 \left\| {{{\bf{D}}_{{{\bf{R}}_k}}}{{\bf{F}}_k}({\bf{\Delta}})} \right\|_2\leq\frac{{\rho ({\bf{H}}_{k,k}^H{{\bf{H}}_{k,k}})}}{{{{\left( {{\lambda _{\min }}\left( {{\bf{H}}_{k,k}^H{{\bf{T}}_k}{{\bf{H}}_{k,k}}} \right)} \right)}^2}}}
\end{equation}

Finally, by combining (\ref{compatibility}) with  (\ref{Jociab}) and (\ref{second}), we have
\begin{equation}\label{inequlityyy}
    {\left\| {{{\bf{F}}_k}({\bf{Q}}_{ - k}^{(1)}) - {{\bf{F}}_k}({\bf{Q}}_{ - k}^{(2)})} \right\|_F} \le {\alpha _k}{\left\| {{\bf{Q}}_{ - k}^{(1)} - {\bf{Q}}_{ - k}^{(2)}} \right\|_F}
\end{equation}
where ${\alpha _k}$ is given in (\ref{defineafak}). Hence, if condition (\ref{uniquecondition}) is satisfied, condition (\ref{lemma2}) in Lemma 3 holds, which completes the proof.

\section{Proof of Theorem 3}\label{proofTheorem3}

Before proving the theorem, we introduce some basic definitions that will be used in our derivations. Given the multiuser mapping  ${\bf{F}}({\bf{Q}})$ defined in (\ref{overallmapping}), we introduce the following block-maximum norm on  ${\mathbb{C}^{KM \times KM}}$, defined as \cite{Ortega2000}
 \begin{equation}\label{block-norm}
    {\left\| {{\bf{F}}({\bf{Q}})} \right\|_{F,{\rm{block}}}} \buildrel \Delta \over = \mathop {\max }\limits_{k \in {\Psi _K}} {\left\| {{{\bf{F}}_k}({{\bf{Q}}_{ - k}})} \right\|_F}.
 \end{equation}
 Let ${\left\| \cdot \right\|_{\infty ,{\rm{vec}}}}$  be the vector maximum norm, defined as \cite{Horn-2012}
 \begin{equation}\label{vecnorm}
   {\left\| {\bf{x}} \right\|_{\infty ,{\rm{vec}}}} \buildrel \Delta \over = \mathop {\max }\limits_{k \in {\Psi _K}} |{x_k}|,{\bf{x}} \in {\mathbb{R}^K},
 \end{equation}
and let ${\left\| \cdot \right\|_{\infty ,{\rm{mat}}}}$  denote the matrix norm induced by  ${\left\| \cdot \right\|_{\infty ,{\rm{vec}}}}$, given by \cite{Horn-2012}:
\begin{equation}\label{matnorm}
    {\left\| {\bf{A}} \right\|_{\infty ,{\rm{mat}}}} \buildrel \Delta \over = \mathop {\max }\limits_k \sum\limits_{r = 1}^K {|{{[{\bf{A}}]}_{k,r}}|} ,{\rm{ }}{\bf{A}} \in {\mathbb{R}^{K \times K}}.
\end{equation}
Finally, we introduce the nonnegative matrix ${\bf{B}} \in \mathbb{R}_ + ^{K \times K}$  defined as
\begin{equation}\label{B}
   {[{\bf{B}}]_{k,r}} \buildrel \Delta \over = \left\{ \begin{array}{l}
{\alpha _k^2},\quad{\rm{ if }}\ k \ne r,\\
0,\quad{\rm{  otherwise}}{\rm{.}}
\end{array} \right.
\end{equation}
Define ${\alpha _{\max }} \buildrel \Delta \over = \mathop {\max }\limits_{k \in {\Psi _K}} {\alpha _k^2}$ and $\beta  \buildrel \Delta \over = (K - 1){\alpha _{\max }}<1$, then ${\left\| {\bf{B}} \right\|_{\infty ,{\rm{mat}}}}$  can be easily computed as
\begin{equation}\label{Bmat}
    {\left\| {\bf{B}} \right\|_{\infty ,{\rm{mat}}}} = \beta.
\end{equation}

Based on the above results, we then give the contraction property of the multiuser mapping function in the following lemma, which will be used in the proof of the theorem.

\itshape \textbf{Lemma 9:}  \upshape If the uniqueness condition in (\ref{uniquecondition}) in Theorem 2 is satisfied, then the multiuser mapping function ${\bf{F}}({\bf{Q}})$  defined in (\ref{overallmapping}) is Lipschitz continuous on  ${\cal W}$:
\begin{equation}\label{block-contraction}
    {\left\| {{\bf{F}}({{\bf{Q}}^{(1)}}) - {\bf{F}}({{\bf{Q}}^{(2)}})} \right\|_{F,{\rm{block}}}} \le \sqrt{ \beta} {\left\| {{{\bf{Q}}^{(1)}} - {{\bf{Q}}^{(2)}}} \right\|_{F,{\rm{block}}}}
\end{equation}
$\forall {{\bf{Q}}^{(1)}},{{\bf{Q}}^{(2)}} \in {\cal W}$, where ${\left\| \cdot \right\|_{F,{\rm{block}}}}$ is defined in (\ref{block-norm}). Furthermore, the mapping is a block-contraction with modulus  $\sqrt{\beta}  < 1$.

\itshape \textbf{Proof:}  \upshape Given  ${{\bf{Q}}^{(1)}} = ({\bf{Q}}_1^{(1)}, \cdots ,{\bf{Q}}_K^{(1)}) \in {\cal W}$ and  ${{\bf{Q}}^{(2)}} = ({\bf{Q}}_1^{(2)}, \cdots ,{\bf{Q}}_K^{(2)}) \in {\cal W}$, define, for each  $k \in {\Psi _K}$,
${e_{{F_k}}} \buildrel \Delta \over = {\left\| {{{\bf{F}}_k}({\bf{Q}}_{ - k}^{(1)}) - {{\bf{F}}_k}({\bf{Q}}_{ - k}^{(2)})} \right\|_F^2}$ and ${e_k} \buildrel \Delta \over = {\left\| {{\bf{Q}}_k^{(1)} - {\bf{Q}}_k^{(2)}} \right\|_F^2}$.
Then we have
\begin{eqnarray}
{e_{{F_k}}}\!\! =\!\!{\left\| {{{\bf{F}}_k}({\bf{Q}}_{ - k}^{(1)})\! -\! {{\bf{F}}_k}({\bf{Q}}_{ - k}^{(2)})} \right\|_F^2}
 \!\!\le\! {\alpha _k^2} {\left\| {{\bf{Q}}_{ - k}^{(1)}\! - \!{\bf{Q}}_{ - k}^{(2)}} \right\|_F}
 \!=\!\! \sum\limits_{j \ne k} {{{[{\bf{B}}]}_{k,j}}{{\left\| {{\bf{Q}}_j^{(1)}\! -\! {\bf{Q}}_j^{(2)}} \right\|}_F^2}}
 \!=\!\! \sum\limits_{j \ne k} {{{[{\bf{B}}]}_{k,j}}{e_j}}\label{inset}
\end{eqnarray}
$\forall {{\bf{Q}}^{(1)}},{{\bf{Q}}^{(2)}} \in {\cal W}$ and  $\forall k \in {\Psi _K}$.

Introducing the vectors	
${{\bf{e}}_F} \buildrel \Delta \over = {[{e_{{F_1}}}, \cdots ,{e_{{F_K}}}]^T},{\rm{and }}\ {\bf{e}} \buildrel \Delta \over = {[{e_1}, \cdots ,{e_K}]^T}$,
the set of inequalities in (\ref{inset}) can be rewritten as:
\begin{equation}\label{rewrinset}
    {\bf{0}} \le {{\bf{e}}_F} \le {\bf{Be}},{\rm{  }}\forall {{\bf{Q}}^{(1)}},{{\bf{Q}}^{(2)}} \in {\cal W}.
\end{equation}
Then we have \cite{Ortega2000}
\begin{equation}\label{inchain}
    {\left\| {{{\bf{e}}_F}} \right\|_{\infty ,{\rm{vec}}}} \le {\left\| {{\bf{Be}}} \right\|_{\infty ,{\rm{vec}}}} \le {\left\| {\bf{B}} \right\|_{\infty ,{\rm{mat}}}}{\left\| {\bf{e}} \right\|_{\infty ,{\rm{vec}}}} = \beta {\left\| {\bf{e}} \right\|_{\infty ,{\rm{vec}}}}
\end{equation}
where ${\left\|  \right\|_{\infty ,{\rm{vec}}}}$  and ${\left\|  \right\|_{\infty ,{\rm{mat}}}}$  are defined in (\ref{vecnorm}) and (\ref{matnorm}), respectively, and the last equality follows from (\ref{Bmat}).

Finally, using (\ref{inchain}) and (\ref{block-norm}), one obtains
\begin{equation}\label{contraction}
\begin{array}{c}
{\left\| {{\bf{F}}({{\bf{Q}}^{(1)}}) - {\bf{F}}({{\bf{Q}}^{(2)}})} \right\|_{F,{\rm{block}}}^2} = {\left\| {{{\bf{e}}_F}} \right\|_{\infty ,{\rm{vec}}}} \le \beta {\left\| {{{\bf{Q}}^{(1)}} - {{\bf{Q}}^{(2)}}} \right\|_{F,{\rm{block}}}^2}
\end{array}
\end{equation}
$\forall {{\bf{Q}}^{(1)}},{{\bf{Q}}^{(2)}} \in {\cal W}$. Hence, the lemma follows.  \hfill $\Box$

Interestingly, we find that if condition (\ref{uniquecondition}) in Theorem 2 is satisfied, the multiuser mapping function ${\bf{F}}({\bf{Q}})$  is not only a block-contraction for the Frobenius norm as shown in Lemma 2 but also block-maximum norm as shown in Lemma 9.

The remaining task is to show that, under condition (\ref{uniquecondition}) in Theorem 2, conditions of the asynchronous convergence theorem in Prop.2.1 of \cite{Bertsekas-1989} on page 431 are satisfied. Based on the above lemma, the proof can be done in a similar fashion to the proof for the rate maximization game in \cite{Scutari-2009} . For this reason, the details are omitted for simplicity.

\section{Proof of Theorem 4}\label{proofTheorem4}
Since  the links are located far away, the interference among the links is negligible so that the covariance matrix of each link $k$ (i.e., ${\bf{R}}_k$) can be regarded as an identity matrix ${\bf{I}}$. Hence, the SE function of links $k$ reduces to
\begin{equation}\label{capacityfunction}
C_k({P_k}) = \mathop {\max }\limits_{{\rm{tr(}}{{\bf{Q}}_k}{\rm{) = }}{P_k}} {\rm{log}}_2\left| {{\bf{I}} + {\bf{H}}_{k,k}^H{{\bf{H}}_{k,k}}{{\bf{Q}}_k}} \right|.
\end{equation}
Obviously, the optimal solution of ${{{\bf{Q}}_k}}$ is ${\bf{Q}}_k^* = {{\bf{\tilde U}}_k}{{\bf{P}}_k}{\bf{\tilde U}}_k^H$, where ${{\bf{\tilde U}}_k}$ is the eigenvector matrix of ${\bf{H}}_{k,k}^H{{\bf{H}}_{k,k}}$, and ${{\bf{P}}_k}$ is nonnegative matrix with ${{\bf{P}}_k} = {\rm{diag}}\{ {p_{k,1}}, \cdots ,{p_{k,{r_k}}}\} $, which is the power allocation over the streams. By inserting ${\bf{Q}}_k^*$ into (\ref{capacityfunction}), the SE function can be equivalently obtained as
\begin{equation}\label{equaliobtain}
\begin{array}{l}
{C_k}({P_k}) = \mathop {{\rm{max}}}\limits_{\{ {p_{k,m}}\} } \sum\limits_{m = 1}^{{r_k}} {{{\log }_2}(1 + {d_{k,m}}{p_{k,m}})} \\
\qquad\qquad{\rm{                s}}{\rm{.t}}{\rm{.  }}\sum\limits_{m = 1}^{{r_k}} {{p_{k,m}}}  \le {P_k}{\rm{,  }}\ {p_{k,m}} \ge {\rm{0}}{\rm{.  }}
\end{array}
\end{equation}
The solution to the above problem can be easily solved as
${p_{k,m}} = {({\mu _k} - d_{k,m}^{ - 1})^ + },m \in \{ 1,2, \cdots ,{r_k}\}$, where ${\mu _k}$ is the water-level chosen to satisfy $\sum\limits_{m = 1}^{{r_k}} {{p_{k,m}}}  = {P_k}$, and ${d_{k,m}}$'s are the eigenvalues of ${{\bf{H}}_{k,k}^H{{\bf{H}}_{k,k}}}$, which are arranged in decreasing order.

Define ${g_l} = ld_{k,l}^{ - 1} - \sum\limits_{m = 1}^l {d_{k,m}^{ - 1}} $ and ${g_{{r_k+1}}} = \infty $. Obviously, ${g_1} = 0$. Then, the SE function can be rewritten as
\begin{equation}\label{rewitecfunc}
{C_k}\left( {{P_k}} \right) = \sum\limits_{m = 1}^l {{{\log }_2}\left( {\frac{{{d_{k,m}}}}{l}\left( {{P_k} + \sum\limits_{m = 1}^l {d_{k,m}^{ - 1}} } \right)} \right)} ,\ {P_k} \in \left[ {{g_l},{g_{l + 1}}}\right], l = 1, \cdots ,{r_k}.
\end{equation}
Note that $l$ represents the number of streams that are allocated with positive power allocation.

Based on the above analysis, we now provide the properties of the SE function in the following lemma.

\itshape \textbf{Lemma 10:}  \upshape The SE function $C_k({P_k})$ is a continuous, strictly increasing, differential and strictly concave function of $P_k$. Moreover, the first derivative of function $C_k({P_k})$ (denoted as ${C_k}^\prime ({P_k})$)  is a continuous function of $P_k$. Also, the SE function $C_k({P_k})$ is twice differentiable in each interval of the subregions $\left[ {{g_l},{g_{l + 1}}} \right],l = 1, \cdots ,{r_k}$, which is given by
\begin{equation}\label{twicediff}
{C''_k}({P_k}) =  - \frac{l}{{\ln 2}}{\left( {{P_k} + \sum\limits_{m = 1}^l {d_{k,m}^{ - 1}} } \right)^{ - 2}} < 0,{P_k} \in ({g_l},{g_{l + 1}}),l = 1, \cdots ,{r_k}.
\end{equation}
\itshape \textbf{Proof:}  \upshape
Obviously, in each interval of the subregions, the SE function $C_k({P_k})$ is a continuous, strictly increasing, differential and concave function of $P_k$. We only need to verify the boundary points.
For the boundary points, we have
\begin{eqnarray}
&&\mathop {\lim }\limits_{{P_k} \to g_{l + 1}^ - } {C_k}({P_k}) = \sum\limits_{m = 1}^l {{{\log }_2}} \left( {\frac{{{d_{k,m}}}}{{{d_{k,l + 1}}}}} \right) = \mathop {\lim }\limits_{{P_k} \to g_{l + 1}^ + } {C_k}({P_k}),\label{conti}\\
&&\mathop {\lim }\limits_{{P_k} \to g_{l + 1}^ - } {C_k}^\prime ({P_k}) = \frac{{{d_{k,l + 1}}}}{{\ln 2}} = \mathop {\lim }\limits_{{P_k} \to g_{l + 1}^ + } {C_k}^\prime ({P_k}) > 0.\label{diff}
\end{eqnarray}
Equation (\ref{conti}) means that the SE function $C_k({P_k})$ is a continuous function and equation (\ref{diff}) means that the  SE function is a strictly increasing and differential function.

Unfortunately, the SE function is not twice differentiable since
\begin{equation}\label{nottwicediff}
 \mathop {\lim }\limits_{{P_k} \to g_{l + 1}^ - } {C_k}^{\prime \prime }({P_k}) =  - \frac{1}{{\ln 2}}\frac{{d_{k,l + 1}^2}}{l} \ne  - \frac{1}{{\ln 2}}\frac{{d_{k,l + 1}^2}}{{l + 1}} = \mathop {\lim }\limits_{{P_k} \to g_{l + 1}^ + } {C_k}^{\prime \prime }({P_k}).
\end{equation}
Hence, the concavity of the SE function cannot be proved by using the above method. To overcome this issue, we use the duality theory to prove the concavity of the SE function. Specifically, for a given $P_k$, the SE function  $C_k({P_k})$ can be obtained by solving problem (\ref{equaliobtain}), which is a strictly convex optimization problem. It can also be verified that the constraints in  problem (\ref{equaliobtain}) satisfy the Slater's condition. Hence, the duality gap for this problem is zero. Thus, the SE function can also be obtained by solving the following min-max problem
\begin{eqnarray}
{C_k}({P_k}) &=& \mathop {\rm{min}}\limits_{\mu  \ge 0} \mathop {\rm{max}}\limits_{{p_{k,m}} \ge 0,\forall m} \sum\limits_{m = 1}^{{r_k}} {{{\log }_2}(1 + {d_{k,m}}{p_{k,m}})}  - \mu (\sum\limits_{m = 1}^{{r_k}} {{p_{k,m}}}  - {P_k})\\
 &=& \mathop {\rm{min}}\limits_{\mu  \ge 0} \sum\limits_{m = 1}^{{r_k}} {{\left[ {{{\log }_2}\left( {\frac{{{d_{k,m}}}}{\mu {\ln 2}}} \right)} \right]^ + }}  - {\sum\limits_{m = 1}^{{r_k}} {\left[ {\frac{1}{{\ln 2}} - \frac{\mu }{{{d_{k,m}}}}} \right]} ^ + } + \mu {P_k}\label{duealpro}\\
 &=& \sum\limits_{m = 1}^{{r_k}} {{\left[ {{{\log }_2}\left( {\frac{{{d_{k,m}}}}{\mu(P_k){\ln 2} }} \right)} \right]^ + }}  - {\sum\limits_{m = 1}^{{r_k}} {\left[ {\frac{1}{{\ln 2}} - \frac{{\mu ({P_k})}}{{{d_{k,m}}}}} \right]} ^ + } + \mu ({P_k}){P_k},
\end{eqnarray}
where $\mu ({P_k}) \ge 0$ denotes the optimal value for given ${P_k}$, which is unique due to the fact that problem (\ref{rewitecfunc}) is a strictly convex problem.

For any two points $P_k^{(1)},P_k^{(2)}$, define $P_k^{(3)} = \theta P_k^{(1)} + (1 - \theta )P_k^{(2)}$, where $\theta  \in (0,1)$. Let $\mu (P_k^{(1)}),\mu (P_k^{(2)})$ and $\mu (P_k^{(3)})$ be the optimal $\mu $ for ${C_k}(P_k^{(1)}),{C_k}(P_k^{(2)})$ and ${C_k}(P_k^{(3)})$, respectively. Then, for $i = 1,2$, we have
\begin{eqnarray}
{C_k}(P_k^i) &=& \sum\limits_{m = 1}^{{r_k}} {{\left[ {{{\log }_2}\left( {\frac{{{d_{k,m}}}}{\mu(P_k^i){\ln 2} }} \right)} \right]^ + }}  - {\sum\limits_{m = 1}^{{r_k}} {\left[ {\frac{1}{{\ln 2}} - \frac{{\mu(P_k^i)}}{{{d_{k,m}}}}} \right]} ^ + } + \mu (P_k^i)P_k^i\\
 &<& \sum\limits_{m = 1}^{{r_k}} {{\left[ {{{\log }_2}\left( {\frac{{{d_{k,m}}}}{\mu(P_k^3){\ln 2}}} \right)} \right]^ + }}  - {\sum\limits_{m = 1}^{{r_k}} {\left[ {\frac{1}{{\ln 2}} - \frac{{\mu (P_k^3)}}{{{d_{k,m}}}}} \right]} ^ + } + \mu (P_k^3)P_k^i,
\end{eqnarray}
where the strict inequality follows due to the facts that ${\mu (P_k^3)}$ is not the optimal solution to problem (\ref{duealpro}) for given $P_k^i,i = 1,2$, and problem (\ref{equaliobtain}) has a unique global solution since it is a strictly convex optimization problem \cite{Boyd2004}. By using the above inequality, we have
\begin{equation}\label{convati}
 \theta {C_k}(P_k^1) + (1 - \theta ){C_k}(P_k^2) < {C_k}(P_k^3) = {C_k}(\theta P_k^1 + (1 - \theta )P_k^2).
\end{equation}
Hence, the SE function is also a strictly concave function of $P_k$.
Obviously, the SE function $C_k({P_k})$ is twice differentiable in each interval of the subregions $\left[ {{g_l},{g_{l + 1}}} \right],l = 1, \cdots ,{r_k}$ and the second-order derivative of $C_k({P_k})$  w.r.t. $P_k$ can be easily calculated in (\ref{twicediff}).   \hfill $\Box$

We now proceed to prove the first part of Theorem 4: the corresponding SE increases with the circuit power consumption. The EE function can be rewritten as
\begin{equation}\label{eepk}
  {\rm{EE}}_k = \frac{{{C_k}({P_k})}}{{{P_k} + {P_{\rm{C}}}}}.
\end{equation}
According to Lemma 10, ${C_k}({P_k})$ is a strictly concave function of ${P_k}$. Moreover, the denominator of the EE function is affine in ${P_k}$. Hence, ${\rm{EE}}_k$ is strictly quasiconcave in ${P_k}$ \cite{Boyd2004}. Then, the optimal solution of $P_k$ (denoted as ${P_k^\star}$) to maximize the EE function is unique \cite{wolfstetter1999topics}  and should satisfy the first order  condition \cite{Boyd2004}: ${{C'}_k}({P_k^\star})({P_k^\star} + {P_{\rm{C}}}) - {C_k}({P_k^\star}) = 0$, which is equivalent to
\begin{equation}\label{expressPc}
   \frac{{{C_k}({P_k^\star}) - {P_k^\star}{{C'}_k}({P_k^\star})}}{{{{C'}_k}({P_k^\star})}} = {P_{\rm{C}}}.
\end{equation}
Our task is to analyze the monotonicity of ${P_k^\star}$ w.r.t. ${P_{\rm{C}}}$. If ${P_k^\star}$ is strictly monotonically increasing w.r.t. ${P_{\rm{C}}}$, the corresponding SE is monotonically increasing w.r.t. ${P_{\rm{C}}}$ as well since the SE is increasing in $P_k$ based on Lemma 10. However, directly proving the monotonicity of ${P_k^\star}$ w.r.t. ${P_{\rm{C}}}$ is difficult. Instead, if the following two conditions hold: 1) ${P_{\rm{C}}}$ is a continuous function of ${P_k^\star}$; 2) there is a one-to-one mapping relation between ${P_k^\star}$ and $P_{\rm{C}}$, this proof can be equivalently transformed into the opposite side  \cite{apostol1974mathematical}:  ${P_{\rm{C}}}$ is strictly monotonically increasing w.r.t. ${P_k^\star}$. Obviously, ${P_{\rm{C}}}$ is a continuous function of ${P_k^\star}$ since both ${C_k}({P_k^\star})$ and ${{C'}_k}({P_k^\star})$ are continuous function of $P_k^\star$ according to Lemma 10. The one-to-one mapping relation between ${P_k^\star}$ and $P_{\rm{C}}$ is also obvious: For given $P_{\rm{C}}$, there is a unique ${P_k^\star}$ satisfying (\ref{expressPc}) since ${\rm{EE}}_k$ is strictly quasiconcave in ${P_k}$ \cite{wolfstetter1999topics}; for given ${P_k^\star}$, the unique ${P_{\rm{C}}}$ can be calculated from the left hand side of (\ref{expressPc}).

The remaining task is to prove that ${P_{\rm{C}}}$ is strictly increasing with ${P_k^\star}$. Define the left hand side of (\ref{expressPc}) as $f({P_k^\star})$. Function $f({P_k^\star})$ is obviously differentiable in each interval of the subregions, i.e., $\left[ {{g_l},{g_{l + 1}}} \right],l = 1, \cdots ,{r_k}$,  which can be calculated as
 \begin{equation}\label{twicdiff}
f'(P_k^*) =  - {{C''}_k}(P_k^*){C_k}(P_k^*){\left( {{C_k}^\prime (P_k^*)} \right)^{ - 2}} > 0,{P_k} \in ({g_l},{g_{l + 1}}),l = 1, \cdots ,{r_k}.
 \end{equation}
where the inequality follows from  the facts that ${C_k}(P_k^*)$ is positive and ${{C''}_k}(P_k^*)$ is negative according to (\ref{twicediff}) in Lemma 10. Combining with the fact that $f({P_k^\star})$ is a continuous function of $P_k^\star$, we conclude that function $f({P_k^\star})$ is a strictly increasing function w.r.t. $P_k^\star$ and the proof for the first part of Theorem 4 completes.

Finally, we prove the last part of Theorem 4: the EE decreases with the circuit power consumption. Supposing $P_{\rm{C}}^{(1)} > P_{\rm{C}}^{(2)}$, define $P_k^{(1)}$ and $P_k^{(2)}$ respectively as the optimal solutions for given $P_{\rm{C}}^{(1)}$ and $P_{\rm{C}}^{(2)}$. Then, we have
\begin{equation}\label{chainihkk}
{\rm{E}}{{\rm{E}}_k}(P_{\rm{C}}^{(1)}) = \frac{{{C_k}(P_k^{(1)})}}{{P_k^{(1)} + P_{\rm{C}}^{(1)}}} < \frac{{{C_k}(P_k^{(1)})}}{{P_k^{(1)} + P_{\rm{C}}^{(2)}}} < \frac{{{C_k}(P_k^{(2)})}}{{P_k^{(2)} + P_{\rm{C}}^{(2)}}} = {\rm{E}}{{\rm{E}}_k}(P_{\rm{C}}^{(2)}),
\end{equation}
where the second inequality follows from the assumption that ${P_k^{(2)}}$ is the optimal solution for given $P_{\rm{C}}^{(2)}$. From the above inequalities, we know that the EE decreases with $P_{\rm{C}}$.
\end{appendices}

\bibliographystyle{IEEEtran}
\vspace{-0.4cm}\bibliography{myre}

\end{document}